\documentclass[twoside]{article}
\usepackage{epsfig}
\usepackage{url}

\newlength{\figurewidth}
\newcommand{\hide}[1]{}

\newcommand{\PostNet}{{Post network}}
\newcommand{\BlogNet}{{Blog network}}
\newcommand{\CGM}{{Cascade generation model}}
\newcommand{\B}{{\beta}}  

\newtheorem{observation}{Observation}

\renewcommand{\sectionmark}[1]{\markboth{\textsc{J. Leskovec et al.}}
{\textsc{The Dynamics of Viral Marketing}}}
\renewcommand{\subsectionmark}[1]{\markboth{\textsc{J. Leskovec et al.}}
{\textsc{The Dynamics of Viral Marketing}}}

\newcommand{\captionfonts}{\small}
\makeatletter
\long\def\@makecaption#1#2{%
  \vskip\abovecaptionskip
  \sbox\@tempboxa{{\captionfonts #1: #2}}%
  \ifdim \wd\@tempboxa >\hsize
    {\captionfonts #1: #2\par}
  \else
    \hbox to\hsize{\hfil\box\@tempboxa\hfil}%
  \fi
  \vskip\belowcaptionskip}
\makeatother

\begin{document}

\title{Cascading Behavior in Large Blog Graphs
\footnote{This work also appears in: Leskovec, J., McGlohon, M.,
Faloutsos, F., Glance, N., Hurst, M. 2007. Cascading Behavior in Large
Blog Graphs. SIAM International Conference on Data Mining (SDM) 2007.}}

\author{
  Jure Leskovec, Mary McGlohon, Christos Faloutsos\\
  \textit{\normalsize Machine Learning Department, Carnegie Mellon University,
  Pittsburgh, PA.}\\
  Natalie Glance, Matthew Hurst\\
  \textit{\normalsize Neilsen Buzzmetrics, Pittsburgh, PA.}
}
\date{}

\maketitle

\begin{abstract}
How do blogs cite and influence each other? How do such links evolve?
Does the popularity of old blog posts drop exponentially with time?
These are some of the questions that we address in this work. Our goal
is to build a model that generates realistic cascades, so that it can
help us with link prediction and outlier detection.

Blogs (weblogs) have become an important medium of information because
of their timely publication, ease of use, and wide availability. In
fact, they often make headlines, by discussing and discovering evidence
about political events and facts. Often blogs link to one another,
creating a publicly available record of how information and influence
spreads through an underlying social network. Aggregating links from
several blog posts creates a directed graph which we analyze to discover
the patterns of information propagation in blogspace, and thereby
understand the underlying social network. Not only are blogs interesting
on their own merit, but our analysis also sheds light on how rumors,
viruses, and ideas propagate over social and computer networks.

Here we report some surprising findings of the blog linking and
information propagation structure, after we analyzed one of the largest
available datasets, with $45,000$ blogs and $\approx 2.2$ million
blog-postings. Our analysis also sheds light on how rumors, viruses, and
ideas propagate over social and computer networks. We also present a
simple model that mimics the spread of information on the blogosphere,
and produces information cascades very similar to those found in real
life.
\end{abstract}

\section{Introduction}
\label{sec:intro} Blogs have become an important medium of communication
and information on the World Wide Web.  Due to their accessible and
timely nature, they are also an intuitive source for data involving the
spread of information and ideas.  By examining linking propagation
patterns from one blog post to another, we can infer answers to some
important questions about the way information spreads through a social
network over the Web.  For instance, does traffic in the network exhibit
bursty, and/or periodic behavior? After a topic becomes popular, how
does interest die off -- linearly, or exponentially?

In addition to temporal aspects, we would also like to discover
topological patterns in information propagation graphs (cascades). We
explore questions like: do graphs of information cascades have common
shapes? What are their properties? What are characteristic in-link
patterns for different nodes in a cascade? What can we say about the
size distribution of cascades?

Finally, how can we build models that generate realistic cascades?

\subsection{Summary of findings and contributions}
{\em Temporal patterns:} For the two months of observation, we found
that blog posts do {\em not} have a bursty behavior; they only have a
weekly periodicity. Most surprisingly, the popularity of posts drops
with a {\em power law}, instead of exponentially, that one may have
expected. Surprisingly, the exponent of the power law is $\approx$-1.5,
agreeing very well with Barabasi's theory of heavy tails in human
behavior~\cite{barabasi05human}.

{\em Patterns in the shapes and sizes of cascades and blogs:} Almost
every metric we measured, followed a power law. The most striking result
is that the size distribution of cascades (= number of involved posts),
follows a perfect Zipfian distribution, that is, a power law with slope
=-2. The other striking discovery was on the shape of cascades. The most
popular shapes were the ``stars'', that is, a single post with several
in-links, but none of the citing posts are themselves cited.

{\em Generating Model:} Finally, we design a flu-like epidemiological
model. Despite its simplicity, it generates cascades that match several
of the above power-law properties of real cascades. This model could be
useful for link prediction, link-spam detection, and ``what-if''
scenarios.

\subsection{Paper organization}

In section~\ref{sec:related} we briefly survey related work. We
introduce basic concepts and terminology in section~\ref{sec:prelim}.
Next, we describe the blog dataset, and discuss the data cleaning steps.
We describe temporal link patterns in section~\ref{sec:observ}, and
continue with exploring the characteristics of the information cascades.
We develop and evaluate the \CGM\ in section~\ref{sec:models}. We
discuss implications of our findings in section~\ref{sec:discussion},
and conclude in section~\ref{sec:conclusion}.

\section{Related work}
\label{sec:related} To our knowledge this work presents the first
analysis of temporal aspects of blog link patterns, and gives detailed
analysis about cascades and information propagation on the blogosphere.
As we explore the methods for modeling such patterns, we will refer to
concepts involving power laws and burstiness, social networks in the
blog domain, and information cascades.

\subsection{Burstiness and power laws}

How often do people create blog posts and links? Extensive work has been
published on patterns relating to human behavior, which often generates
bursty traffic. Disk accesses, network traffic, web-server traffic all
exhibit burstiness. Wang et al in~\cite{Wang02Data} provide fast
algorithms for modeling such burstiness. Burstiness is often related to
self-similarity, which was studied in the context of World Wide Web
traffic~\cite{Crovella96Self}. Vazquez et al \cite{Vazquez:2006}
demonstrate the bursty behavior in web page visits and corresponding
response times.

Self-similarity is often a result of heavy-tailed dynamics. Human
interactions may be modeled with networks, and attributes of these
networks often follow \emph{power law}
distributions~\cite{faloutsos99powerlaw}. Such distributions have a PDF
(probability density function) of the form $p(x) \propto x^\gamma$,
where $p(x)$ is the probability to encounter value $x$ and $\gamma$ is
the exponent  of the power law. In log-log scales, such a PDF gives a
straight line with slope $\gamma$. For $\gamma < -1$, we can show that
the Complementary Cumulative Distribution Function (CCDF) is also a
power law with slope $\gamma + 1$, and so is the rank-frequency plot
pioneered by Zipf~\cite{Zipf49Human}, with slope $1/(1+\gamma)$. For
$\gamma = -2$ we have the standard Zipf distribution, and for other
values of $\gamma$ we have the generalized Zipf distribution.

Human activity also follows periodicities, like daily, weekly and yearly
periodicities, often in combination with the burstiness.

\subsection{Blogs}
Most work on modeling link behavior in large-scale on-line data has been
done in the blog domain~\cite{Lada05election,Adar:2005,Kumar:2003}. The
authors note that, while information propagates between blogs, examples
of genuine cascading behavior appeared relatively rare. This may,
however, be due in part to the Web-crawling and text analysis techniques
used to infer relationships among posts~\cite{Adar:2005,Gruhl:2004a}.
Our work here differs in a way that we concentrate solely on the
propagation of links, and do not infer additional links from text of the
post, which gives us more accurate information.

There are several potential models to capture the structure of the
blogosphere.  Work on information diffusion based on
topics~\cite{Gruhl:2004a} showed that for some topics, their popularity
remains constant in time (``chatter'') while for other topics the
popularity is more volatile (``spikes''). Authors in~\cite{Kumar:2003}
analyze community-level behavior as inferred from blog-rolls --
permanent links between ``friend'' blogs. Analysis based on thresholding
as well as alternative probabilistic models of node activation is
considered in the context of finding the most influential nodes in a
network~\cite{Kempe:2003}, and for viral
marketing~\cite{Richardson:2002}. Such analytical work posits a known
network, and uses the model to find the most influential nodes; in the
current work we observe real cascades, characterize them, and build
generative models for them.

\subsection{Information cascades and epidemiology}

Information cascades are phenomena in which an action or idea becomes
widely adopted due to the influence of others, typically, neighbors in
some network~\cite{Bikchandani:1992,Goldenberg:2001,Granovetter:1978}.
Cascades on random graphs using a threshold model have been
theoretically analyzed~\cite{Watts:2002}. Empirical analysis of the
topological patterns of cascades in the context of a large product
recommendation network is in~\cite{jurij05patterns}
and~\cite{jure06viral}.

The study of epidemics offers powerful models for analyzing the spread
of viruses. Our topic propagation model is based on the \emph{SIS}
(Susceptible-Infected-Susceptible) model of
epidemics~\cite{Bailey1975Diseases}. This is models flu-like viruses,
where an entity begin as ``susceptible'', may become ``infected'' and
infectious, and then heals to become susceptible again. A key parameter
is the infection probability $\beta$, that is, the probability of a
disease transmission in a single contact. Of high interest is the {\em
epidemic threshold}, that is, the critical value of $\beta$, above which
the virus will spread and create an epidemic, as opposed to becoming
extinct. There is a huge literature on the study of epidemics on full
cliques, homogeneous graphs, infinite graphs (see~\cite{Hethcote:2000}
for a survey), with recent studies on power-law
networks~\cite{Equiluz02Epidemic} and arbitrary
networks~\cite{WangCWF03}.

\section{Preliminaries}

\label{sec:prelim}
\begin{figure*}
  \begin{center}
  \begin{tabular}{c c c}
    \epsfig{file=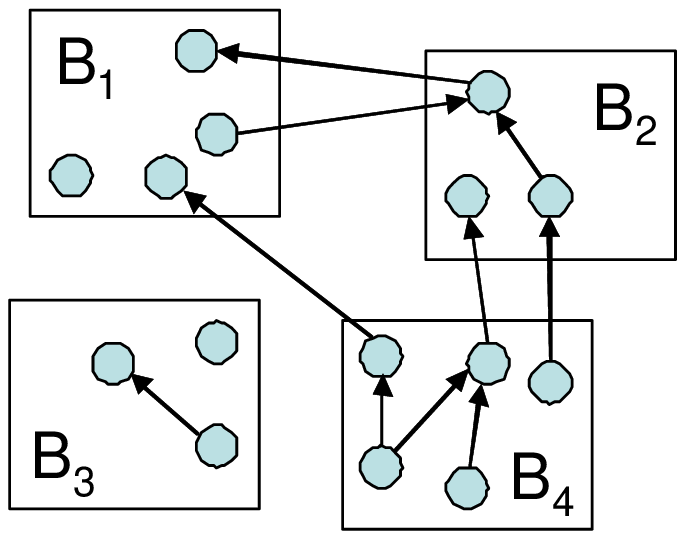, width=0.30\textwidth} &
    \epsfig{file=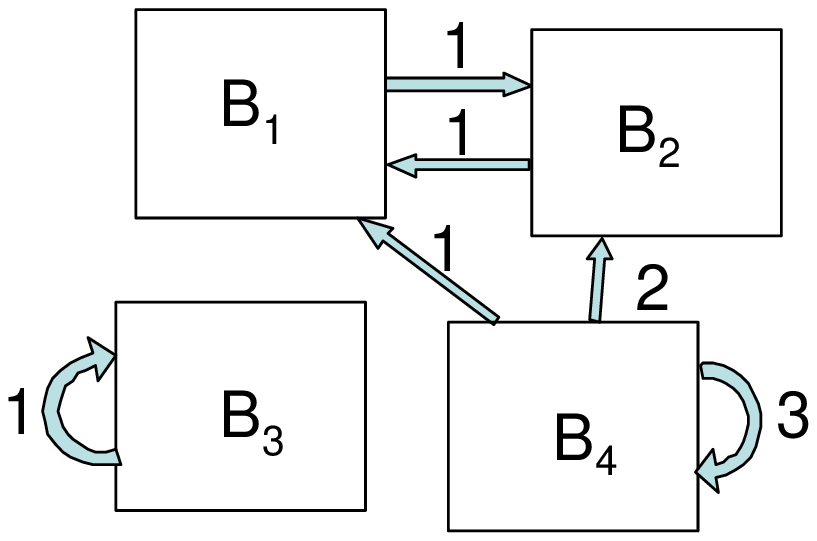, width=0.35\textwidth} &
    \epsfig{file=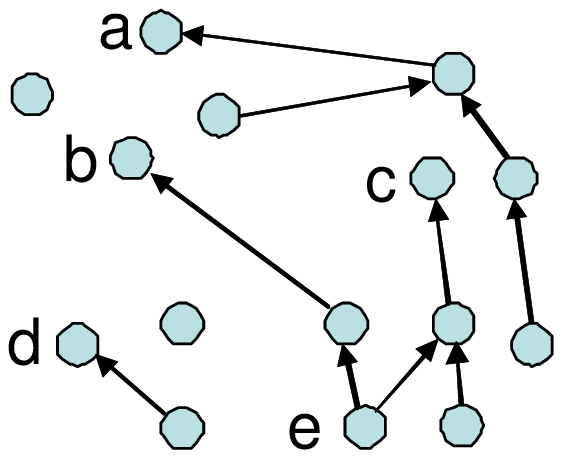, width=0.27\textwidth} \\
    (a)  Blogosphere & (b) \BlogNet\ & (c) \PostNet\ \\
  \end{tabular}
  \end{center}
  \caption{The model of the blogosphere (a). Squares represent blogs and
  circles blog-posts. Each post belongs to a blog, and can contain
  hyper-links to other posts and resources on the web. We create two
  networks: a weighted blog network (b) and a post network (c). Nodes
  $a, b, c,d$ are {\em cascade initiators}, and node $e$ is a {\em
  connector}.}
  \label{fig:blogosphere}
\end{figure*}

In this section we introduce terminology and basic concepts regarding
the blogosphere and information cascades.

Blogs (weblogs) are web sites that are updated on a regular basis. Blogs
have the advantage of being easy to access and update, and have come to
serve a variety of purposes. Often times individuals use them for online
diaries and social networking, other times news sites have blogs for
timely stories. Blogs are composed of posts that typically have room for
comments by readers -- this gives rise to discussion and opinion forums
that are not possible in the mass media.  Also, blogs and posts
typically link each other, as well as other resources on the Web. Thus,
blogs have become an important means of transmitting information. The
influence of blogs was particularly relevant in the 2004 U.S. election,
as they became sources for campaign fundraising as well as an important
supplement to the mainstream media~\cite{Lada05election}. The
blogosphere has continued to expand its influence, so understanding the
ways in which information is transmitted among blogs is important to
developing concepts of present-day communication.

We model two graph structures emergent from links in the blogosphere,
which we call the \emph{\BlogNet} and the \emph{\PostNet}.
Figure~\ref{fig:blogosphere} illustrates these structures. Blogosphere
is composed of blogs, which are further composed of posts. Posts then
contain links to other posts and resources on the web.

From Blogosphere (a), we obtain the \BlogNet\ (b) by collapsing all
links between blog posts into weighted edges between blogs.  A directed
blog-to-blog edge is weighted with the total number of links occurring
between posts in source blog pointing to posts in destination blog.
From the \BlogNet\ we can infer a social network structure, under the
assumption that blogs that are ``friends'' link each other often.

In contrast, to obtain the \PostNet\ (c), we ignore the posts' parent
blogs and focus on the link structure.  Associated with each post is
also the time of the post, so we label the edges in \PostNet\ with the
time difference $\Delta$ between the source and the destination posts.
Let $t_u$ and $t_v$ denote post times of posts $u$ and $v$, where $u$
links to $v$, then the link time $\Delta = t_u - t_v$. Note $\Delta>0$,
since a post can not link into the future and there are no self-edges.

\begin{figure}
  \centering
  \epsfig{file=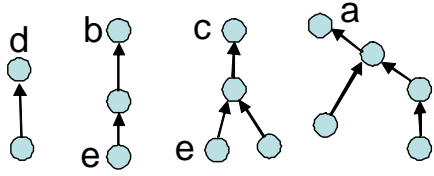, width=0.45\textwidth}
  \caption{Cascades extracted from Figure~\ref{fig:blogosphere}.
  Cascades represent the flow of information through nodes in the
  network.  To extract a cascade we begin with an initiator with no
  out-links to other posts, then add nodes with edges linking to the
  initiator, and subsequently nodes that link to any other nodes in the
  cascade.}
  \label{fig:cascade}
\end{figure}

From the \PostNet, we extract information cascades, which are induced
subgraphs by edges representing the flow of information. A cascade (also
known as conversation tree) has a single starting post called the {\em
cascade initiator} with no out-links to other posts (e.g. nodes
$a,b,c,d$ in Figure~\ref{fig:blogosphere}(c)). Posts then join the
cascade by linking to the initiator, and subsequently new posts join by
linking to members within the cascade, where the links obey time order
($\Delta>0$). Figure~\ref{fig:cascade} gives a list of cascades
extracted from \PostNet\ in Figure~\ref{fig:blogosphere}(c). Since a
link points from the follow-up post to the existing (older) post,
influence propagates following the reverse direction of the edges.

We also define a \emph{non-trivial} cascade to be a cascade containing
at least two posts, and therefore a \emph{trivial cascade} is an
isolated post. Figure~\ref{fig:cascade} shows all non-trivial cascades
in Figure~\ref{fig:blogosphere}(c), but not the two trivial cascades.
Cascades form two main shapes, which we will refer to as \emph{stars}
and \emph{chains}.  A star occurs when a single center posts is linked
by several other posts, but the links do not propagate further. This
produces a wide, shallow tree. Conversely, a chain occurs when a root is
linked by a single post, which in turn is linked by another post.  This
creates a deep tree that has little breadth. As we will later see most
cascades are somewhere between these two extreme points. Occasionally
separate cascades might be joined by a single post -- for instance, a
post may summarize a set of topics, or focus on a certain topic and
provide links to different sources that are members of independent
cascades. The post merging the cascades is called a \emph{connector
node}. Node $e$ in Figure~\ref{fig:cascade}(c) is a connector node. It
appears in two cascades by connecting cascades starting at nodes $b$ and
$c$.

\section{Experimental setup}
\label{sec:data}
\subsection{Dataset description}

We extracted our dataset from a larger set which contains 21.3 million
posts from 2.5 million blogs from August and September
2005~\cite{GlanceHNSST05}. Our goal here is to study temporal and
topological characteristics of information propagation on the
blogosphere. This means we are interested in blogs and posts that
actively participate in discussions, so we biased our dataset towards
the more active part of the blogosphere.

We collected our dataset using the following procedure. We started with
a list of the most-cited blog posts in August 2005. For all posts we
traversed the full conversation tree forward and backward following
post's in- and out-links. For practical reasons we limited the depth of
such conversation trees to 100 and the maximum number of links followed
from a single post to 500. This process gave us a set of posts
participating in conversations. From the posts we extracted a list of
all blogs. This gave us a set of about $45,000$ active blogs. Now, we
went back to the original dataset and extracted all posts coming from
this set of active blogs.

This process produced a dataset of $2,422,704$ posts from $44,362$ blogs
gathered over a two-month period from beginning of August to end of
September 2005. There are the total of $4,970,687$ links in the dataset
out of which $245,404$ are among the posts of our dataset and the rest
point to other resources (e.g. images, press, news, web-pages). For each
post in the dataset we have the following information: unique Post ID,
the URL of the parent blog, Permalink of the post, Date of the post,
post content (html), and a list of all links that occur in the post's
content. Notice these posts are not a random sample of all posts over
the two month period but rather a set of posts biased towards active
blogs participating in conversations (by linking to other posts/blogs).

\begin{figure}
  \centering
  \epsfig{file=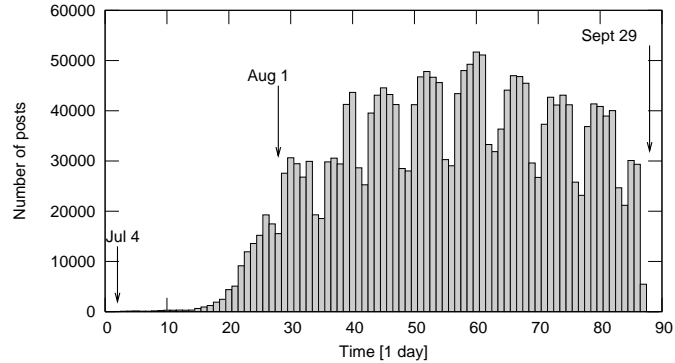, width=0.70\textwidth}
  \caption{Number of posts by day over the three-month period.}
  \label{fig:postovertime}
\end{figure}

In Figure~\ref{fig:postovertime} we plot the number of posts per day
over the span of our dataset. The periodicities in traffic on a weekly
basis will be discussed in section~\ref{sec:observ}. Notice that our
dataset has no ``missing past'' problem, i.e. the starting points of
conversation are not missing due to the beginning of data collection,
since we followed the conversation all the way to its starting point and
thus obtained complete conversations.
The posts span the period from July to September 2005 (90 days), while
the majority of the data comes from August and September. The July posts
in the dataset are parts of conversations that were still active in
August and September.

\subsection{Data preparation and cleaning}

We represent the data as a cluster graph where clusters correspond to
blogs, nodes in the cluster are posts from the blog, and hyper-links
between posts in the dataset are represented as directed edges. Before
analysis, we cleaned the data to most clearly represent the structures
of interest.

\textbf{Only consider out-links to posts in the dataset.} We removed
links that point to posts outside our dataset or other resources on the
web (images, movies, other web-pages). The major reason for this is that
we only have time-stamps for the posts in the dataset while we know
nothing about creation time of URLs outside the dataset, and thus we
cannot consider these links in our temporal analysis.

\textbf{Use time resolution of one day.}  While posts in blogspace are
often labeled with complete time-stamps, many posts in our dataset do
not have a specific time stamp but only the date is known. Additionally,
there are challenges in using time stamps to analyze emergent behaviors
on an hourly basis, because posts are written in different time zones,
and we do not normalize for this. Using a coarser resolution of one day
serves to reduce the time zone effects. Thus, in our analysis the time
differences are aggregated into 24-hour bins.

\textbf{Remove edges pointing into the future.}  Since a post cannot
link to another post that has not yet been written, we remove all edges
pointing into the future. The cause may be human error, post update, an
intentional back-post, or time zone effects; in any case, such links do
not represent information diffusion.

\textbf{Remove self edges.} Again, self edges do not represent
information diffusion. However, we do allow a post to link to another
post in the same blog.

\section{Observations, patterns and laws}
\label{sec:observ}

\subsection{Temporal dynamics of posts and links}

\begin{figure}[tb]
  \begin{center}
  \begin{tabular}{c c}
    \epsfig{file=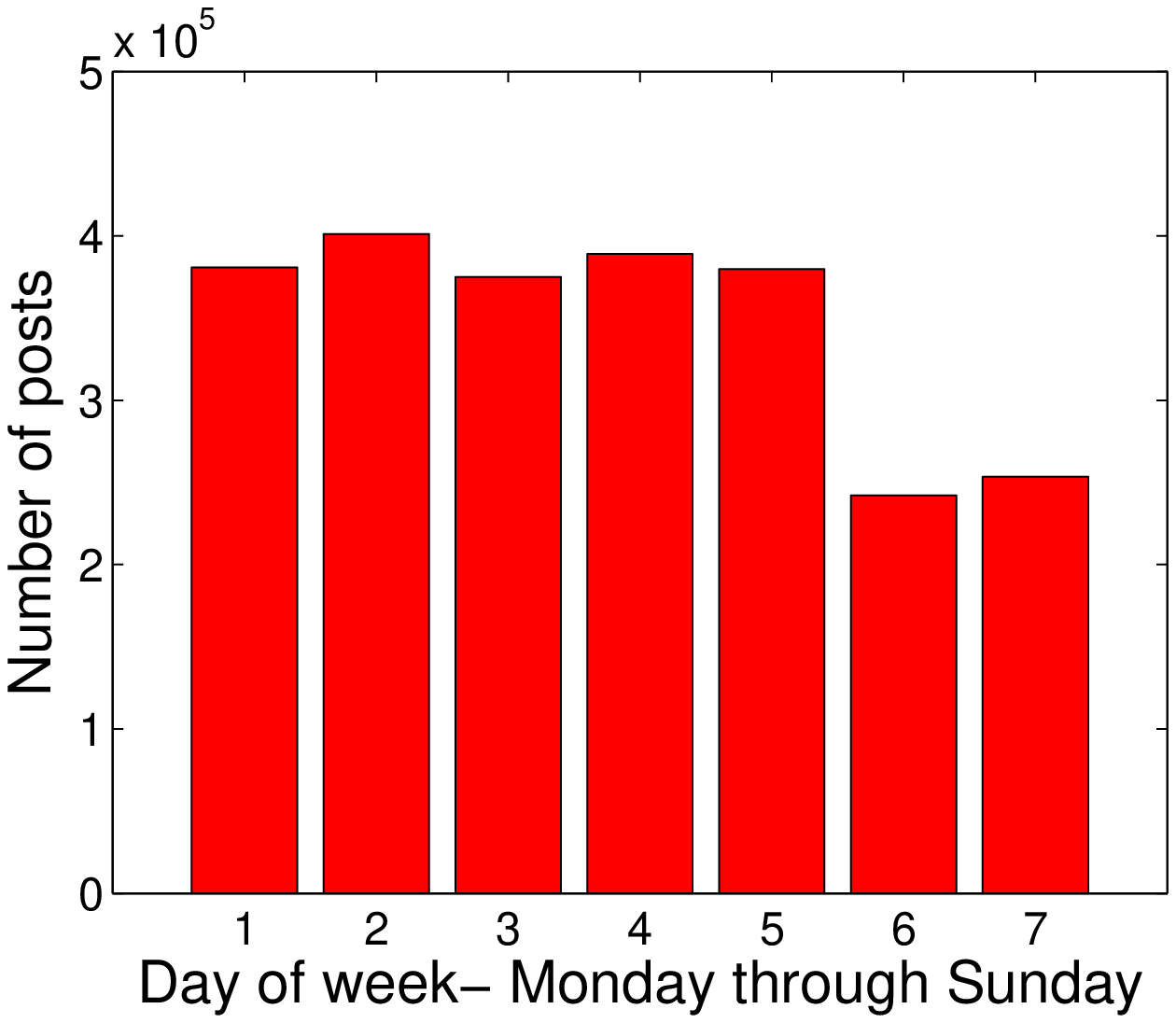, width=0.45\textwidth} &
    \epsfig{file=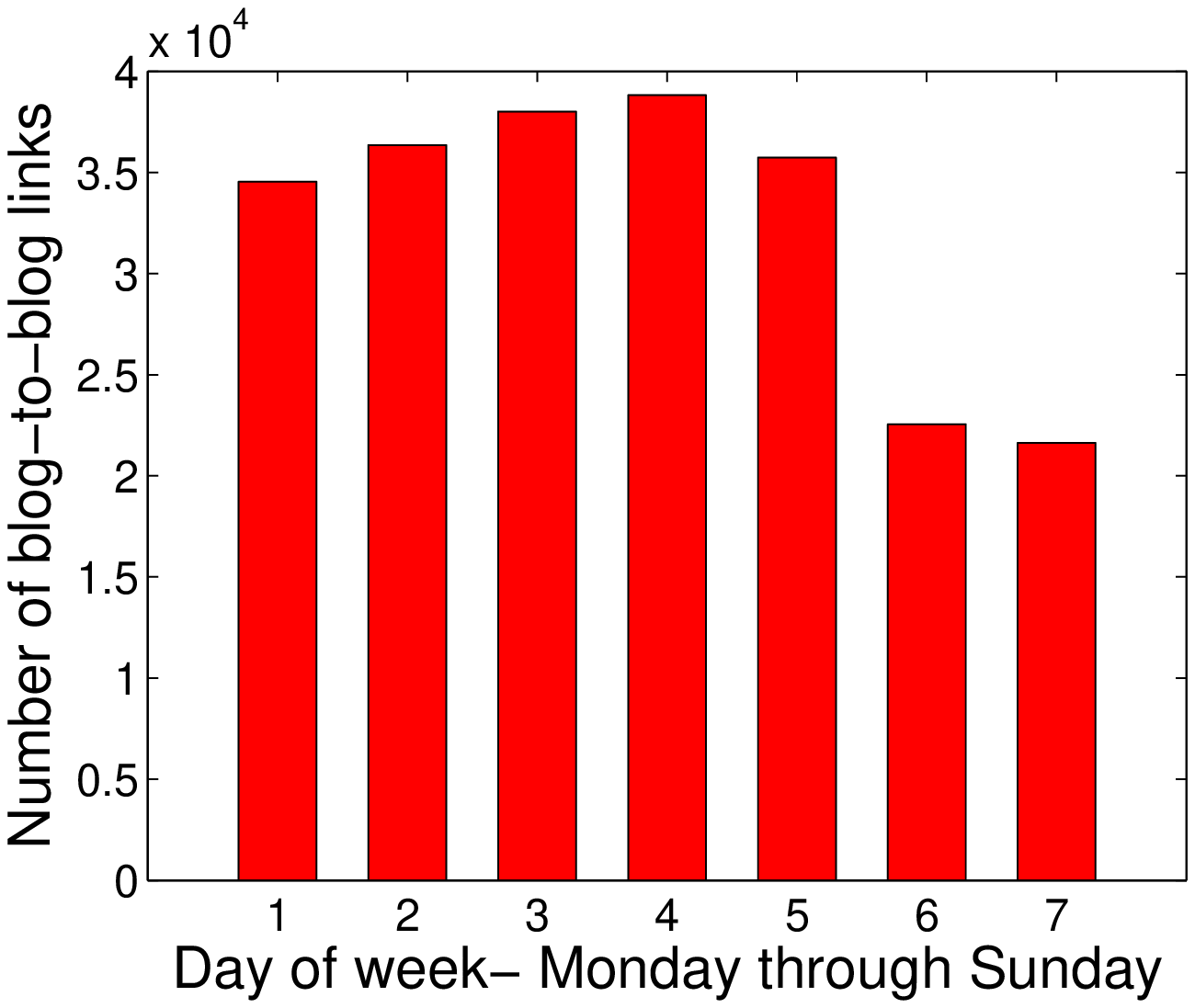, width=0.45\textwidth} \\
    (a) Posts & (b) Blog-to-Blog links\\
  \end{tabular}
  \end{center}
  \caption{Activity counts (number of posts and number of links)
  per day of week, from Monday to Sunday, summed over entire dataset.}
  \label{fig:weekaggregated}
\end{figure}

Traffic in blogosphere is not uniform; therefore, we consider traffic
patterns when analyzing influence in the temporal sense. As
Figure~\ref{fig:postovertime} illustrates, there is a seven-day
periodicity.  Further exploring the weekly patterns,
Figure~\ref{fig:weekaggregated} shows the number of posts and the number
of blog-to-blog links for different days of the week, aggregated over
the entire dataset. Posting and blog-to-blog linking patterns tend to
have a \emph{weekend effect} of sharply dropping off at weekends.

Next, we examine how a post's popularity grows and declines over time.
We collect all in-links to a post and plot the number of links occurring
after each day following the post.  This creates a curve that indicates
the rise and fall of popularity.  By aggregating over a large set of
posts we obtain a more general pattern.

\begin{figure}[tb]
  \begin{center}
  \begin{tabular}{c c}
    \epsfig{file=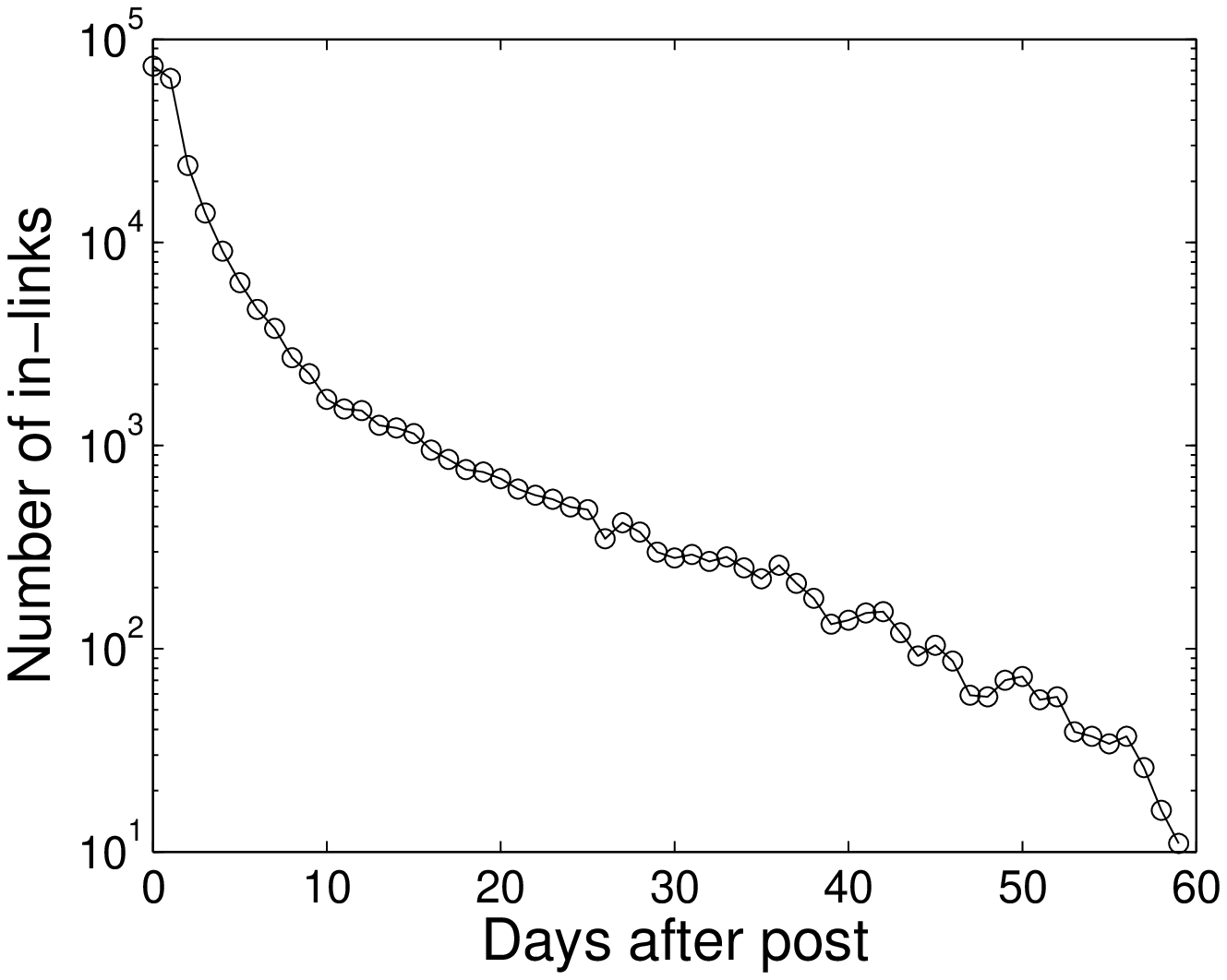, width=0.45\textwidth} &
    \epsfig{file=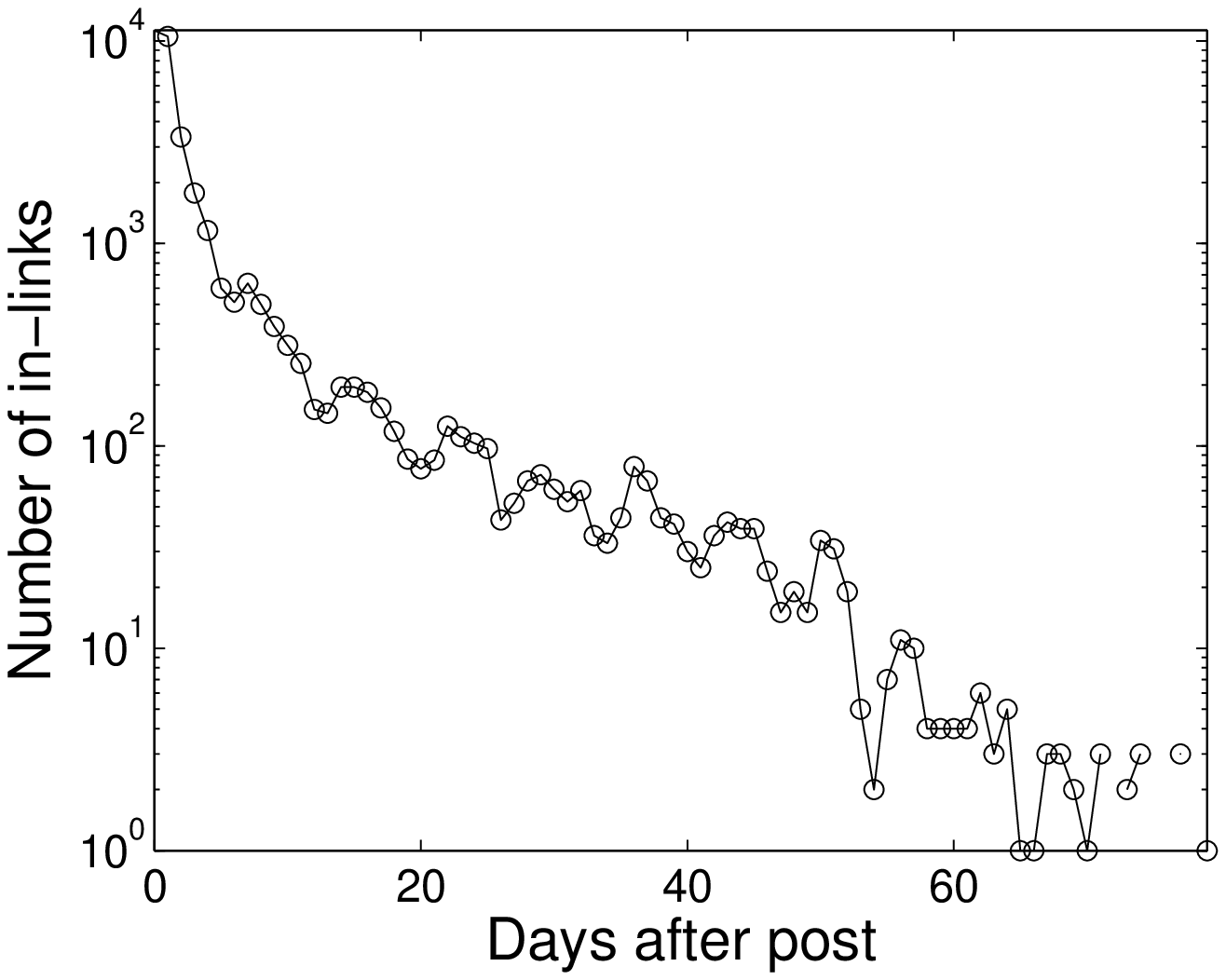, width=0.45\textwidth} \\
    \epsfig{file=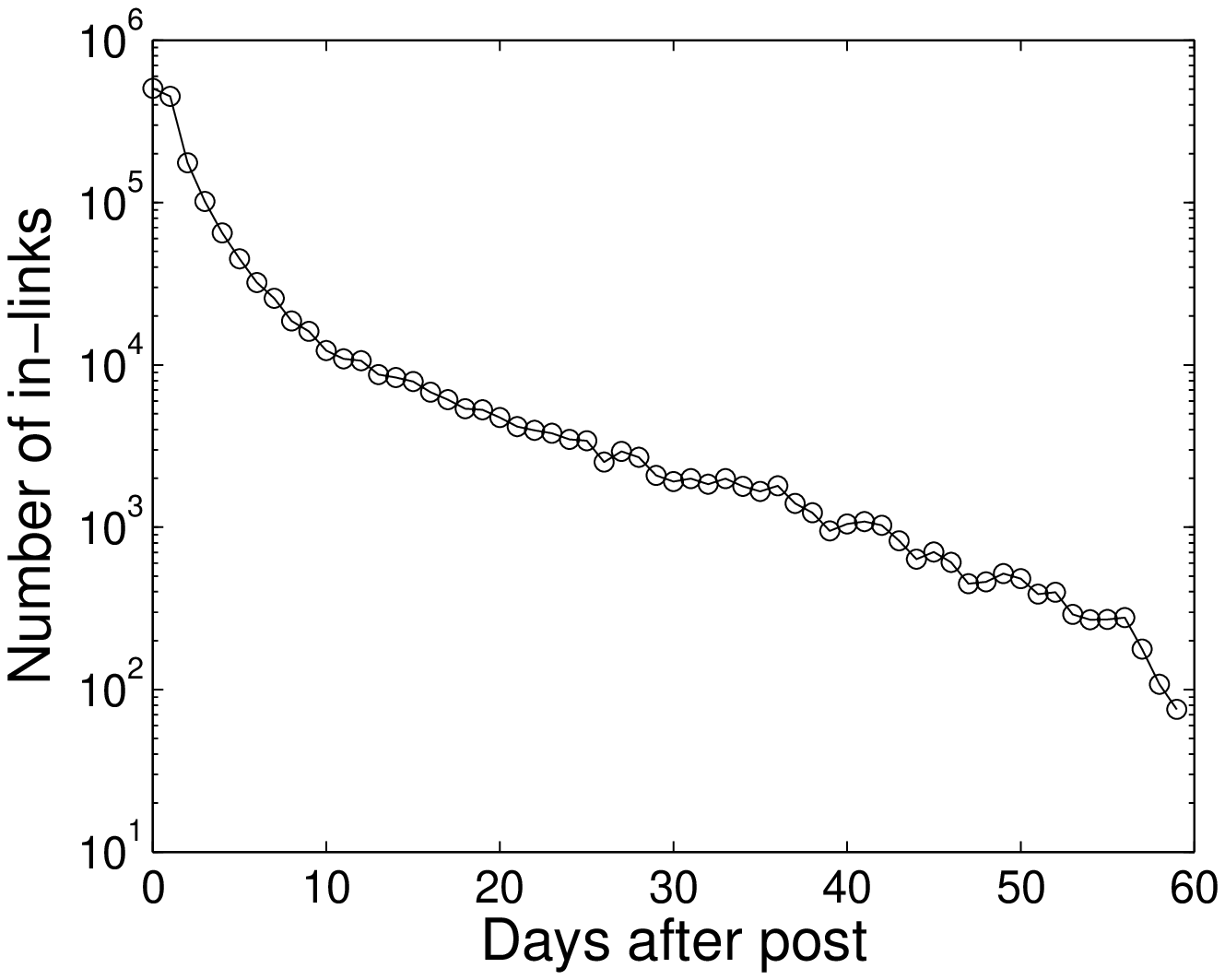, width=0.45\textwidth} &
    \epsfig{file=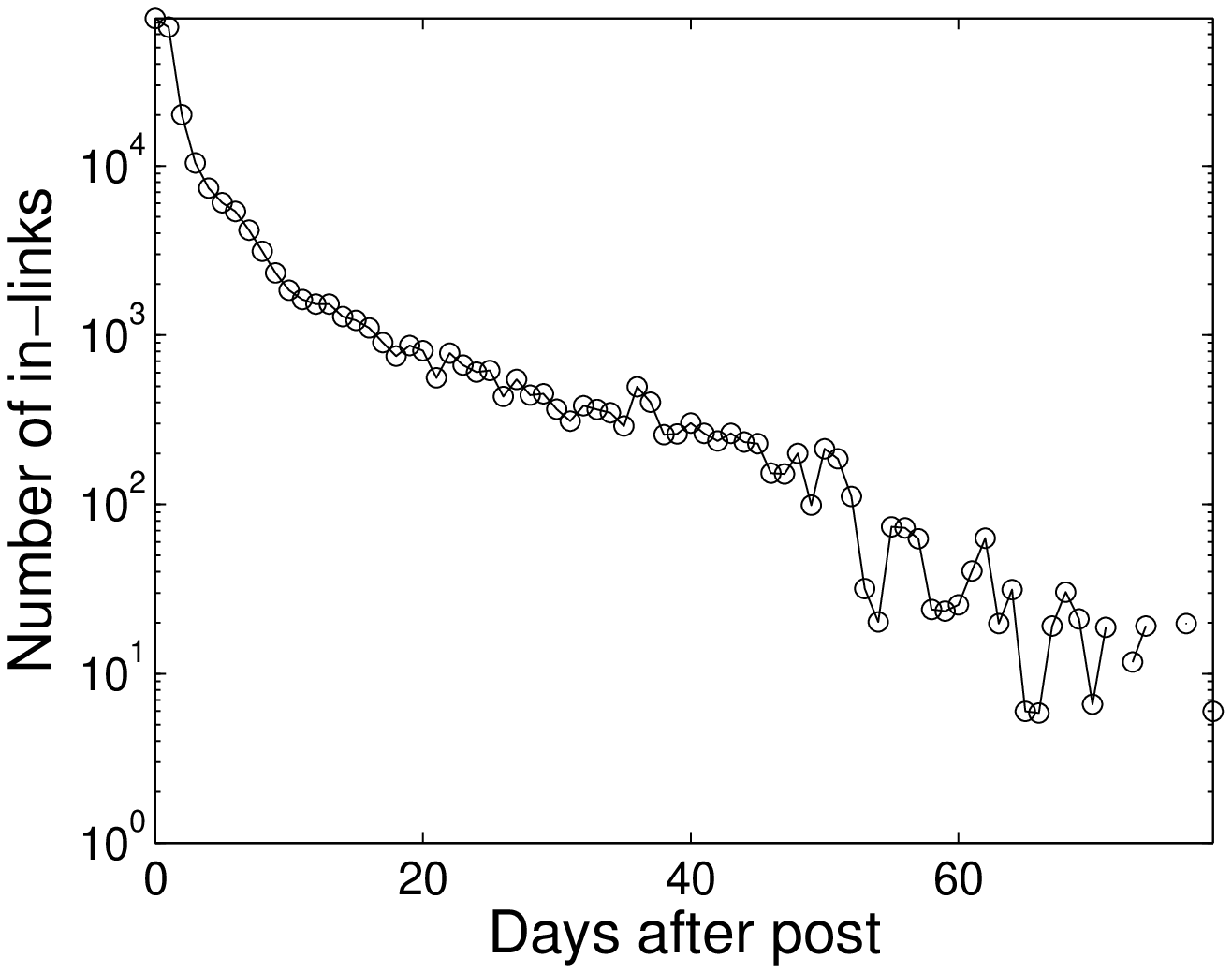, width=0.45\textwidth} \\
    \epsfig{file=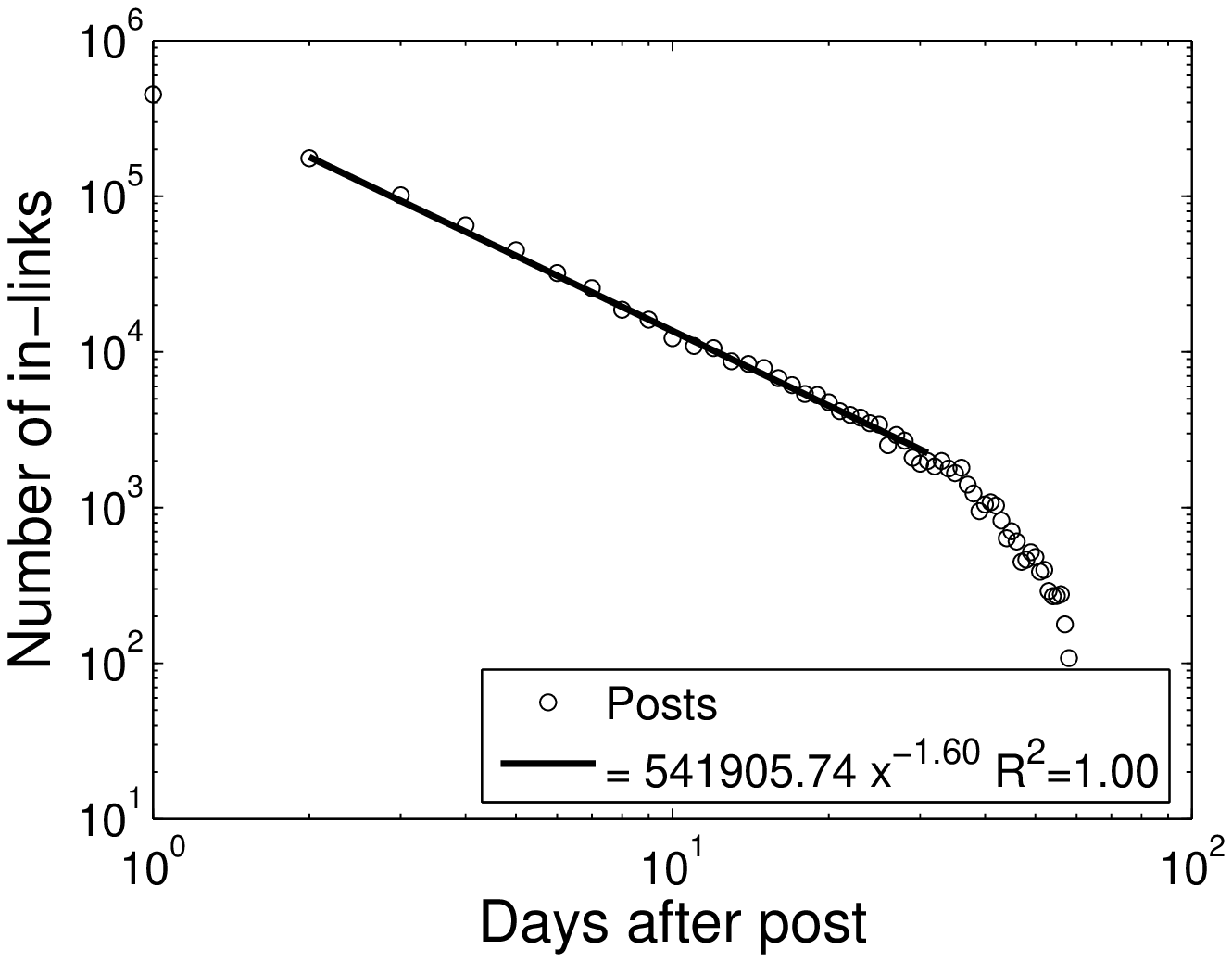, width=0.45\textwidth} &
    \epsfig{file=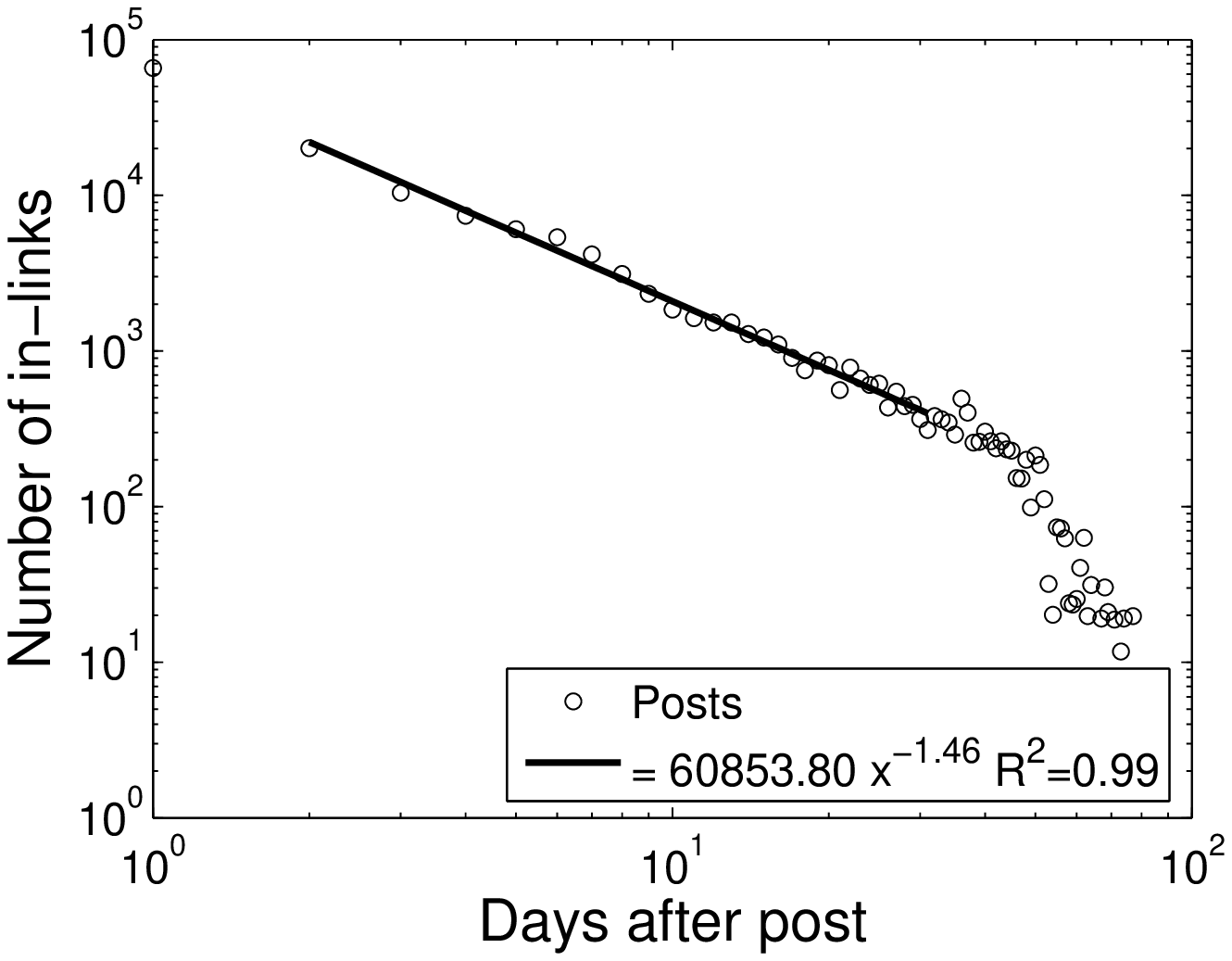, width=0.45\textwidth} \\
    All posts & Only Monday posts \\
  \end{tabular}
  \end{center}
  \caption{Number of in-links vs. the days after the post in
  log-linear scale; when considering all posts (top left), only
  Monday posts (top right). After removing the day-of-the week
  effects (middle row). Power law fit to the data with exponents
  $-1.6$ and $-1.46$ (bottom row).}
  \label{fig:popularity}
\end{figure}

Top left plot of Figure~\ref{fig:popularity} shows number of in-links
for each day following a post for all posts in the dataset, while top
right plot shows the in-link patterns for Monday posts only (in order to
isolate the weekly periodicity).  It is clear that the most links occur
on the first 24 hours after the post, after that the popularity
generally declines. However, in the top right plot, we note that there
are ``spikes'' occurring every seven days, each following Monday.  It
almost appears as if there is compensatory behavior for the sparse
weekend links. However, this is not the case.  Mondays do not have an
unusual number of links; Monday only appears to spike on these graphs
because the natural drop-off of popularity in the following days allows
Monday to tower above its followers.

Thus, fitting a general model to the drop-off graphs may be problematic,
since we might obtain vastly different parameters across posts simply
because they occur at different times during the week. Therefore, we
smooth the in-link plots by applying a weighting parameter to the plots
separated by day of week.  For each delay $\Delta$ on the horizontal
axis, we estimate the corresponding day of week $d$, and we prorate the
count for $\Delta$ by dividing it by $l(d)$, where $l(d)$ is the percent
of blog links occurring on day of week $d$.

This weighting scheme normalizes the curve such that days of the week
with less traffic are bumped up further to meet high traffic days,
simulating a popularity drop-off that might occur if posting and linking
behavior were uniform throughout the week.  A smoothed version of the
post drop-offs is shown in the middle row of
Figure~\ref{fig:popularity}.

We fit the power-law distribution with a cut-off in the tail (bottom
row). We fit on 30 days of data, since most posts in the graph have
complete in-link patterns for the 30 days following publication. We
performed the fitting over all posts and for all days of the week
separately, and found a stable power-law exponent of around $-1.5$,
which is exactly the value predicted by the model where the bursty
nature of human behavior is a consequence of a decision based queuing
process~\cite{barabasi05human} -- when individuals execute tasks based
on some perceived priority, the timing of the tasks is heavy tailed,
with most tasks being rapidly executed, whereas a few experience very
long waiting times.

\begin{observation}
The probability that a post written at time $t_p$ acquires a link at
time $t_p + \Delta$ is:
\[
  p(t_p + \Delta) \propto \Delta^{-1.5}
\]
\end{observation}

\begin{figure}[tb]
  \begin{center}
  \begin{tabular}{c c c}
    \epsfig{file=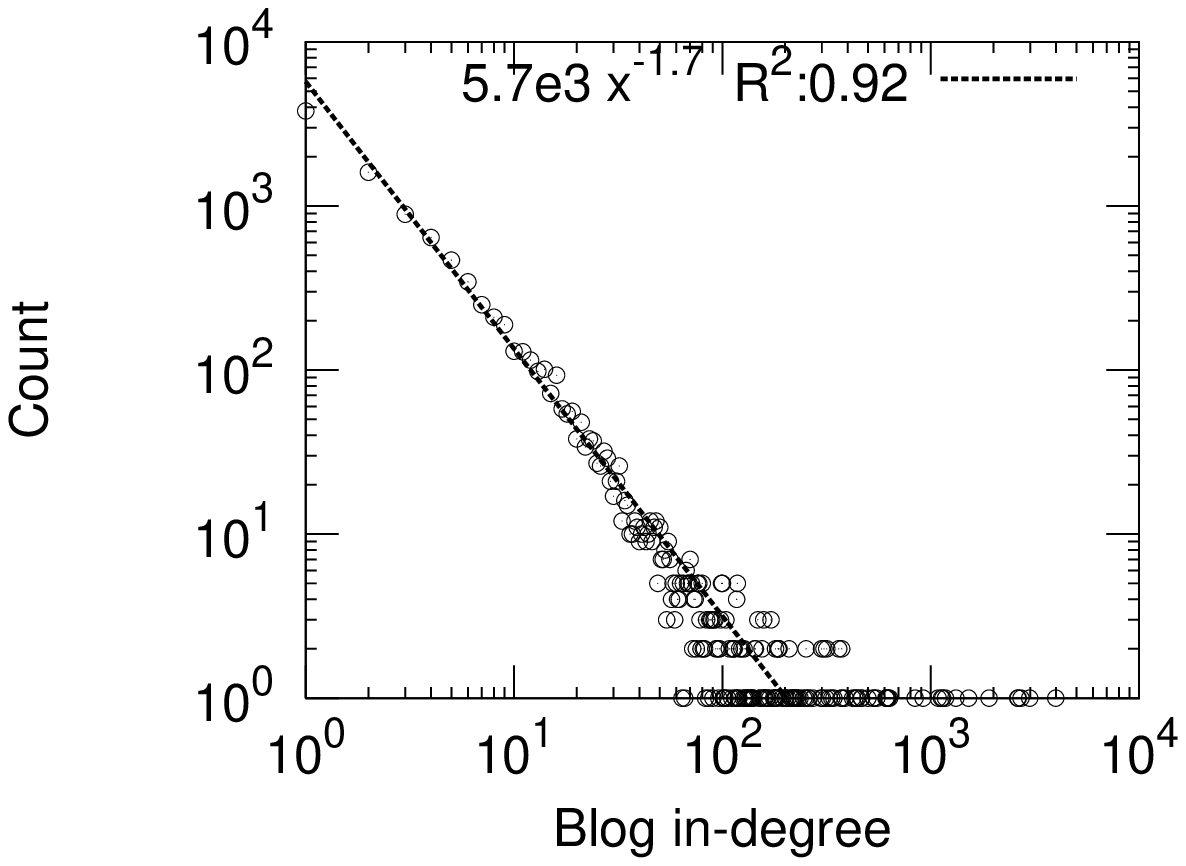, width=0.32\textwidth} &
    \epsfig{file=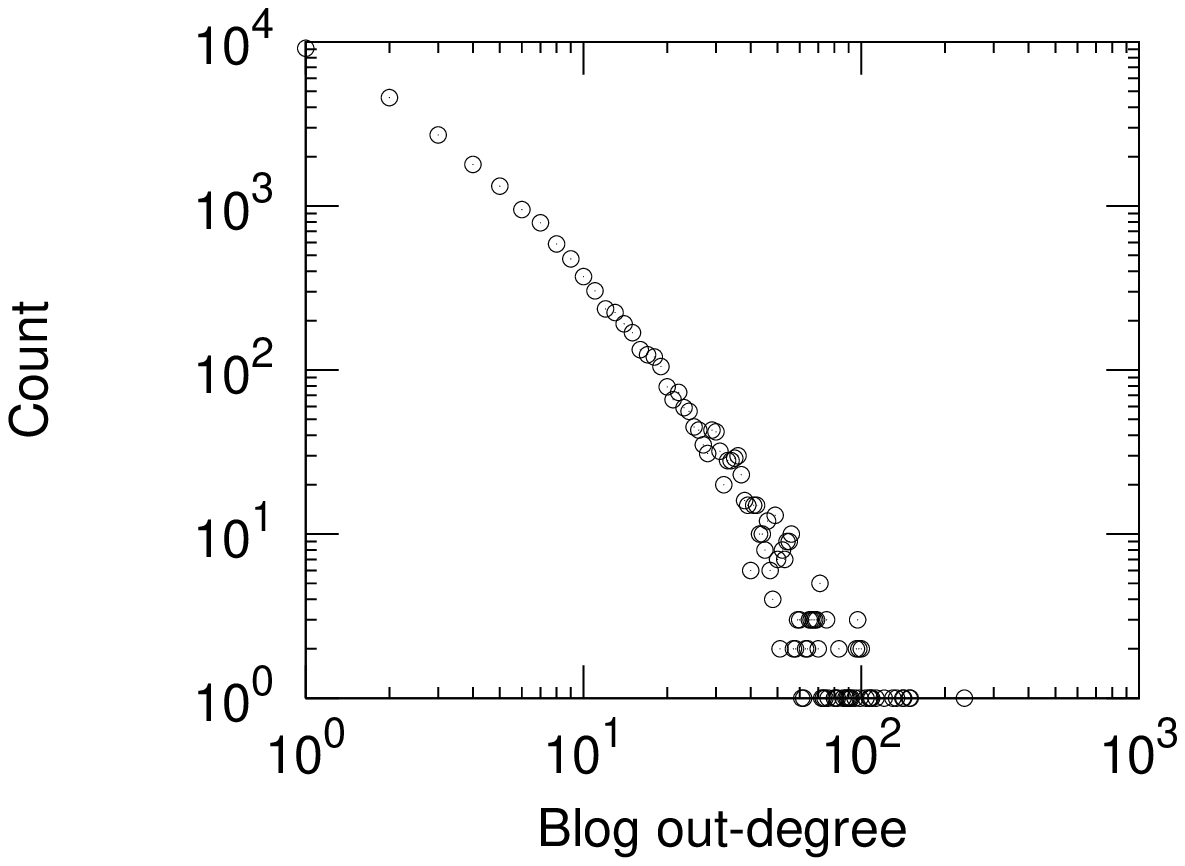, width=0.32\textwidth} &
    \epsfig{file=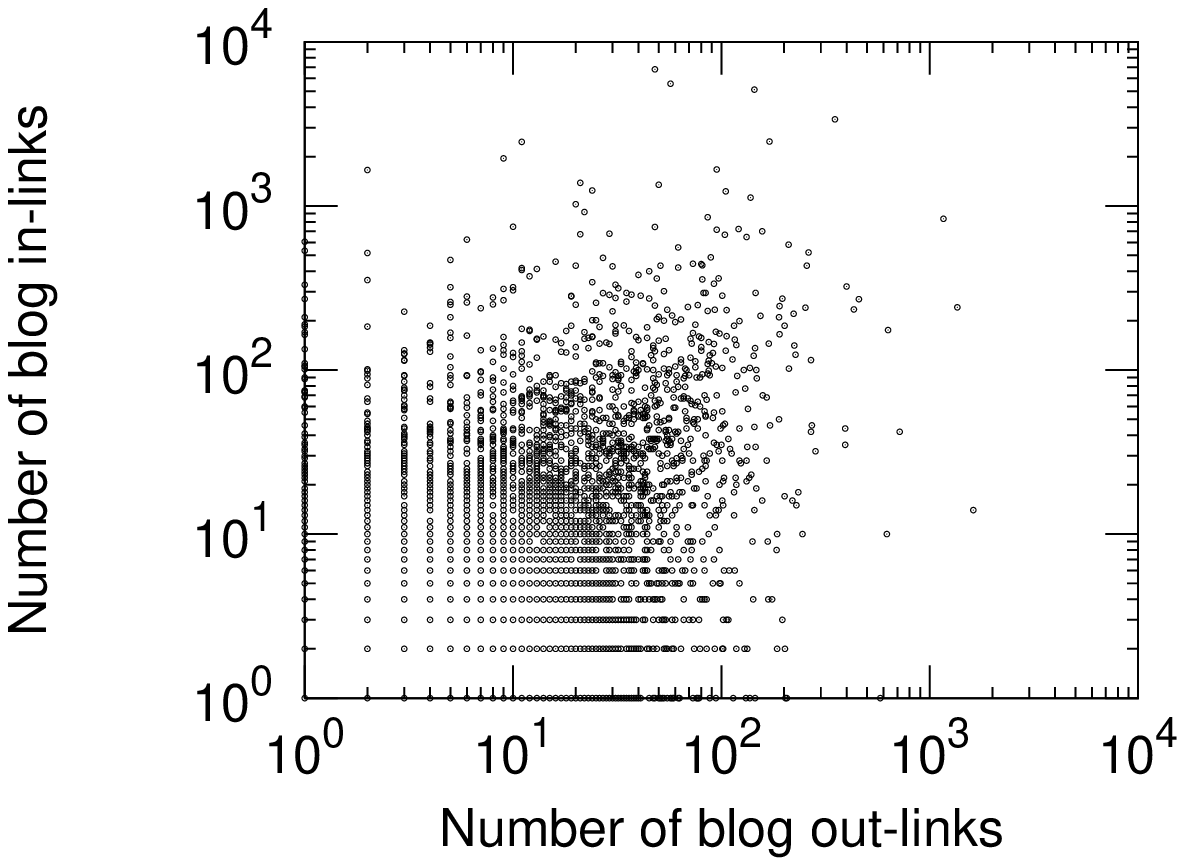, width=0.32\textwidth} \\
    (a) & (b) & (c) \\
  \end{tabular}
  \end{center}
  \caption{(a, b) In- and out-degree distributions of the \BlogNet; (c) the
  scatter plot of the number of in- and out-links of the blogs.}
  \label{fig:indegree}
\end{figure}

\subsection{\BlogNet\ topology}

The first graph we consider is the \BlogNet. As illustrated in
Figure~\ref{fig:blogosphere}(c), every node represents a blog and there
is a weighted directed edge between blogs $u$ and $v$, where the weight
of the edge corresponds to the number of posts from blog $u$ linking to
posts at blog $v$. The network contains $44,356$ nodes and $122,153$
edges. The sum of all edge weights is the number of all post to post
links ($245,404$). Connectivity-wise, half of the blogs belong to the
largest connected component and the other half are isolated blogs.

We show the in- and out-degree distribution in
Figure~\ref{fig:indegree}. Notice they both follow a heavy-tailed
distribution. The in-degree distribution has a very shallow power-law
exponent of $-1.7$, which suggests strong rich-get-richer phenomena. One
would expect that popular active blogs that receive lots of in-links
also sprout many out-links. Intuitively, the attention (number of
in-links) a blog gets should be correlated with its activity (number of
out-links). This does not seem to be the case. The correlation
coefficient between blog's number of in- and out-links is only $0.16$,
and the scatter plot in Figure~\ref{fig:indegree} suggests the same.

\begin{figure}[t]
  \begin{center}
  \begin{tabular}{c c}
    \epsfig{file=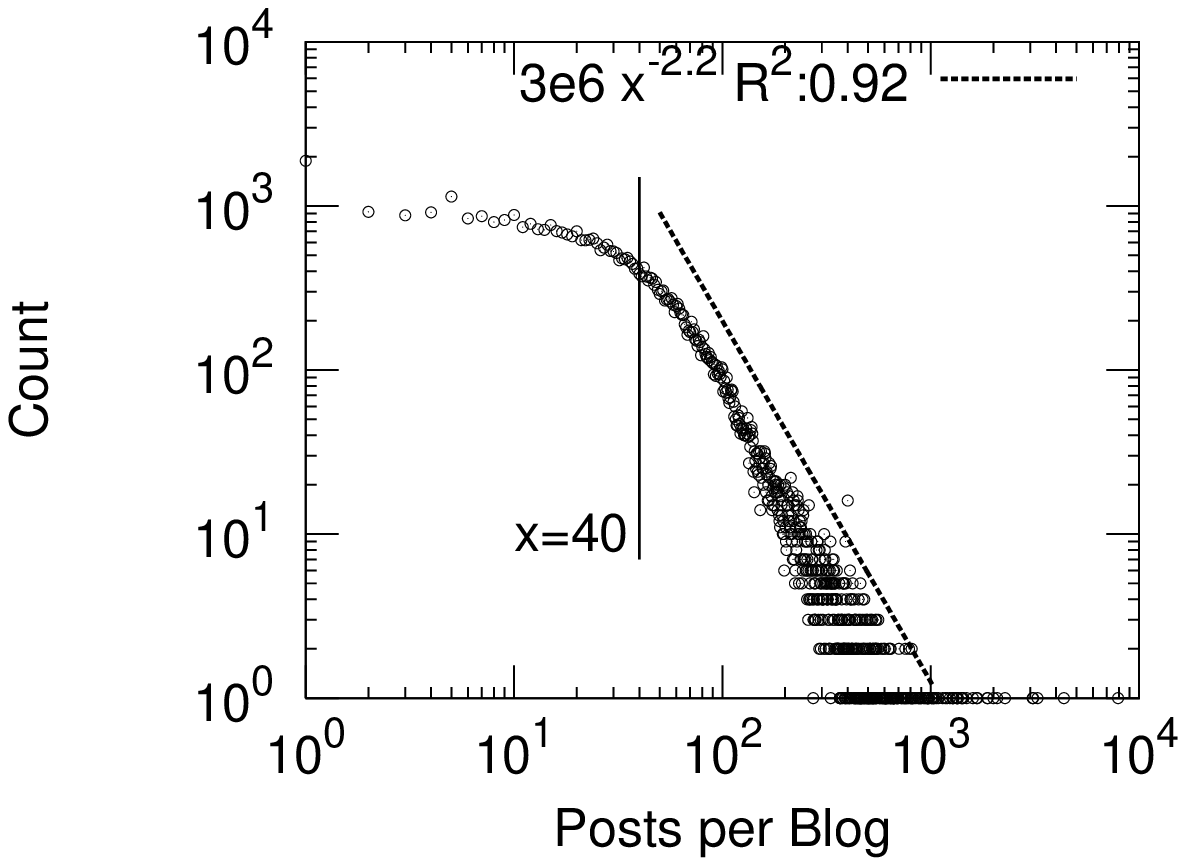, width=0.45\textwidth} &
    \epsfig{file=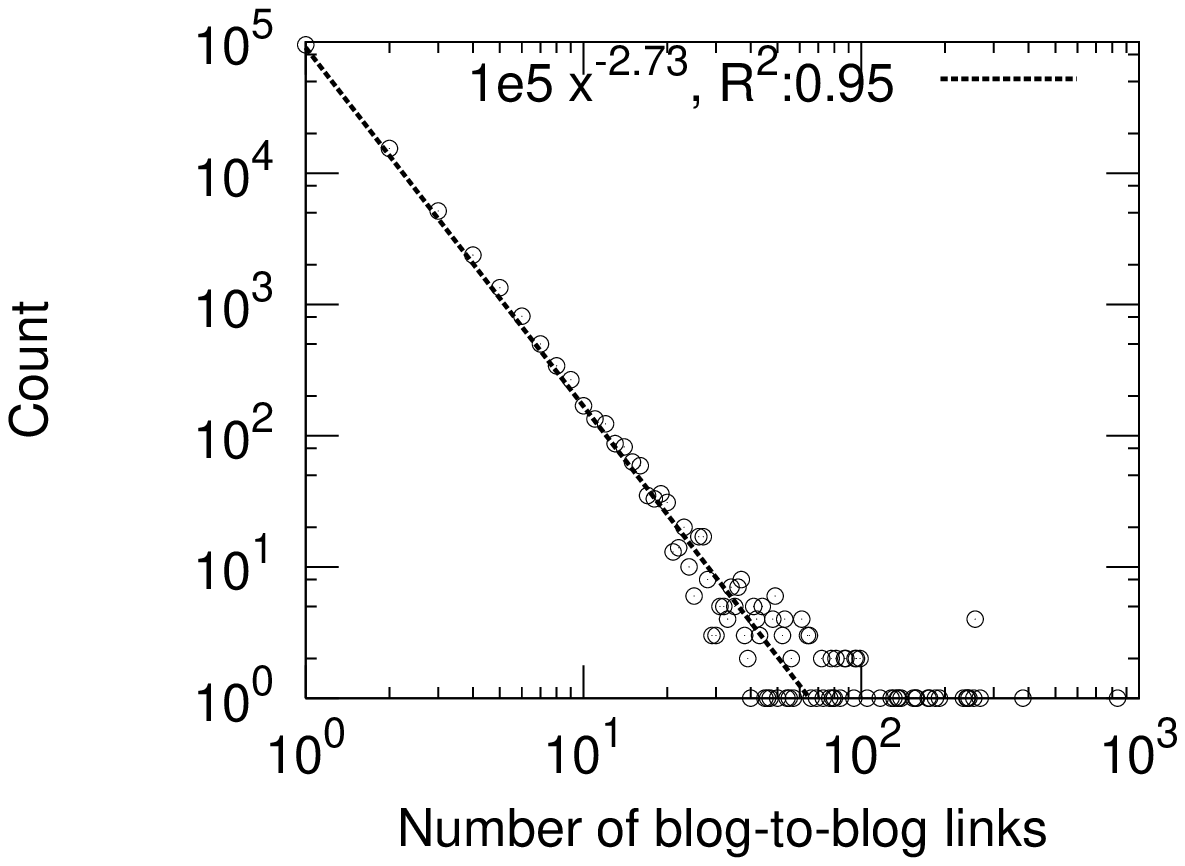, width=0.45\textwidth} \\
  \end{tabular}
  \end{center}
  \caption{Distribution of the number of posts per blog (a);
  Distribution of the number of blog-to-blog links, i.e. the
  distribution over the \BlogNet\ edge weights (b).}
  \label{fig:postsDistr}
\end{figure}

The number of posts per blog, as shown in
Figure~\ref{fig:postsDistr}(a), follows a heavy-tailed distribution. The
deficit of blogs with low number of posts and the knee at around 40
posts per blog can be explained by the fact that we are using a dataset
biased towards active blogs. However, our biased sample of the blogs
still maintains the power law in the number of blog-to-blog links (edge
weights of the \BlogNet) as shown in \ref{fig:postsDistr}(b). The
power-law exponent is $-2.7$.

\begin{figure}[t]
  \begin{center}
  \begin{tabular}{c c}
    \epsfig{file=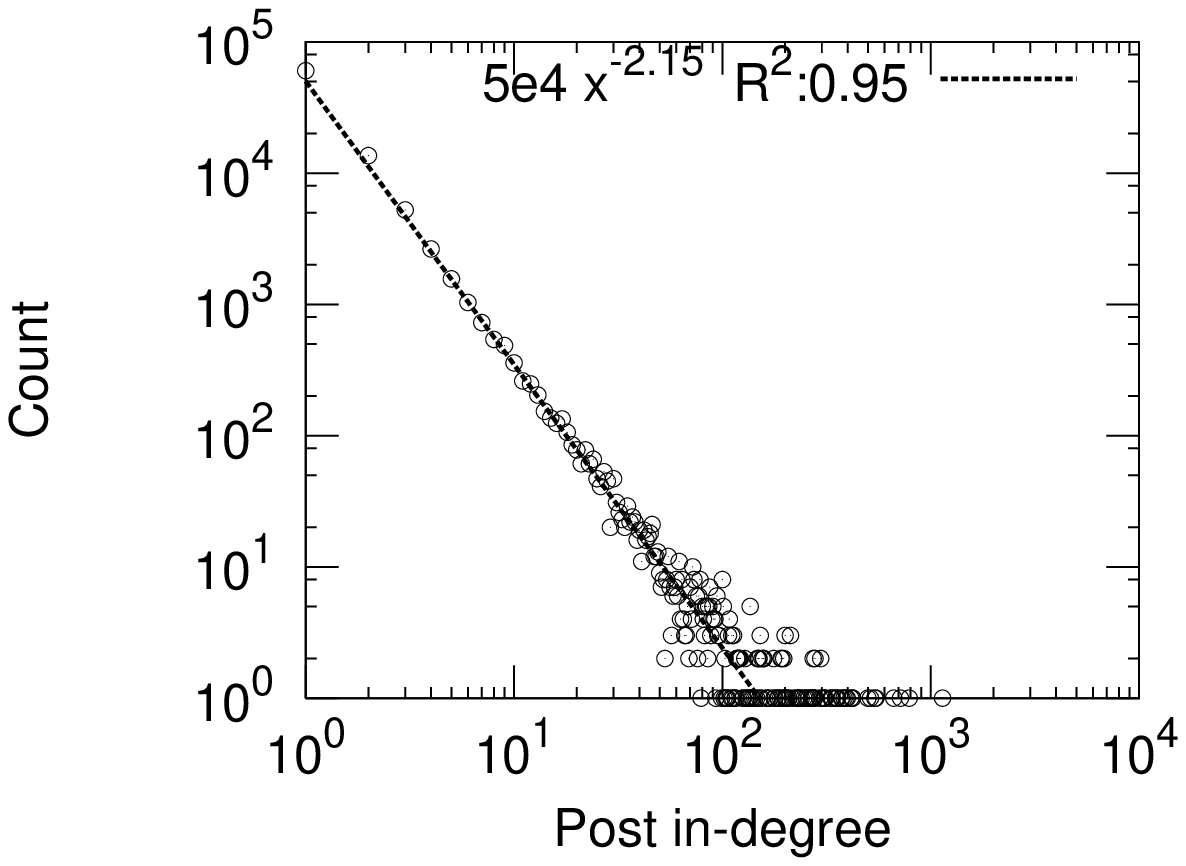, width=0.45\textwidth} &
    \epsfig{file=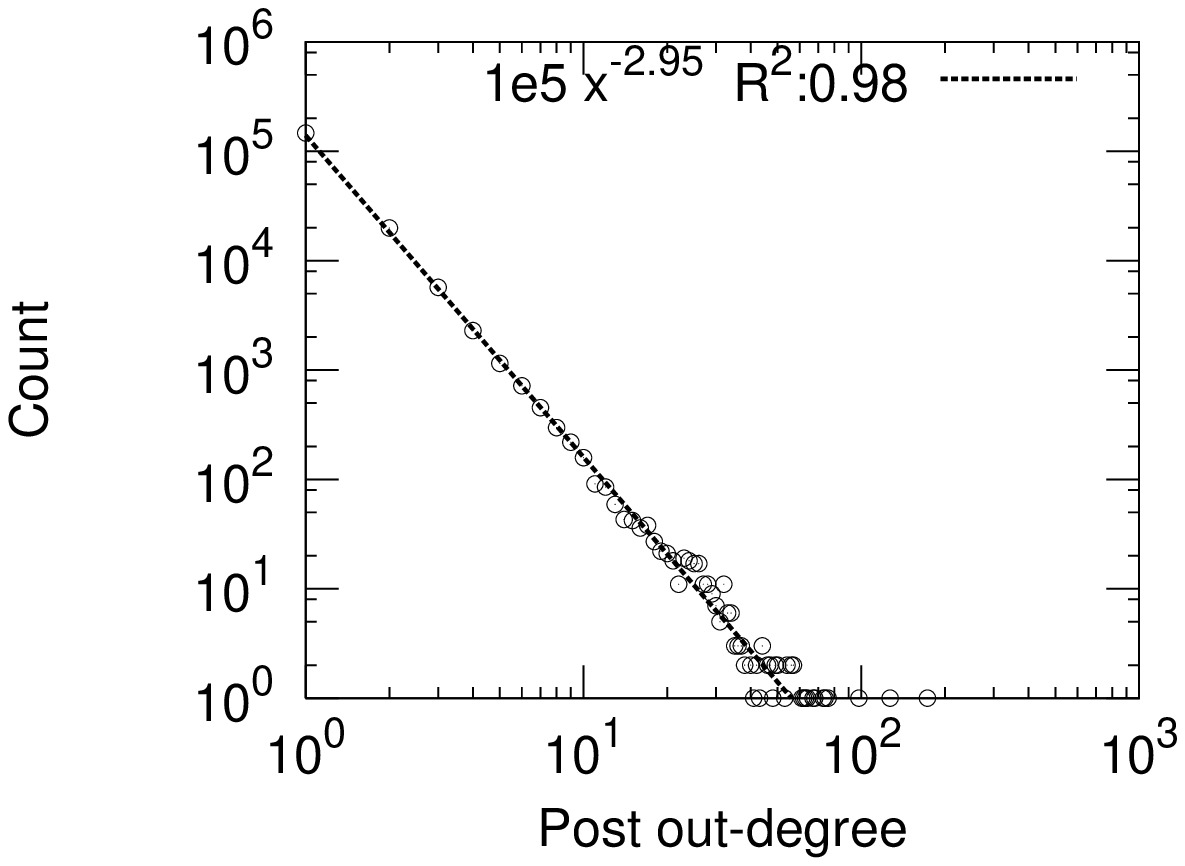, width=0.45\textwidth} \\
  \end{tabular}
  \end{center}
  \caption{\PostNet\ in- and out-degree distribution.}
  \label{fig:postDeg}
\end{figure}

\begin{figure*}[t]
  \centering
  \begin{tabular}{ccccccccccc}
  \includegraphics[scale=0.20, angle=180]{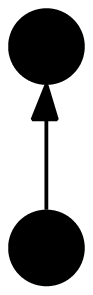} &
  \includegraphics[scale=0.20, angle=180]{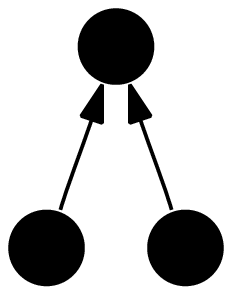} &
  \includegraphics[scale=0.20, angle=180]{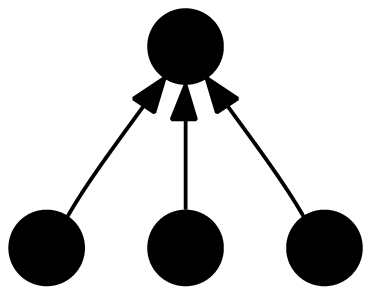} &
  \includegraphics[scale=0.20, angle=180]{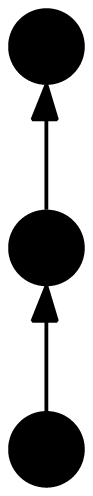} &
  \includegraphics[scale=0.20, angle=180]{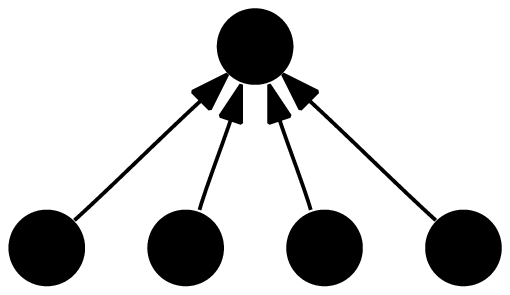} &
  \includegraphics[scale=0.20, angle=180]{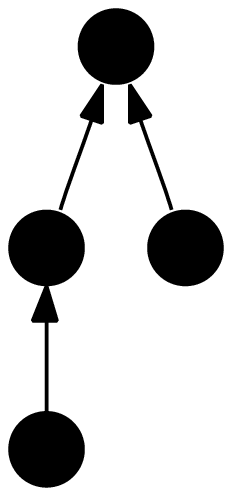} &
  \includegraphics[scale=0.20, angle=180]{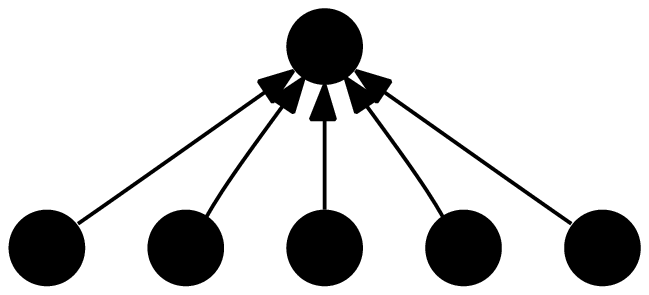} &
  \includegraphics[scale=0.20, angle=180]{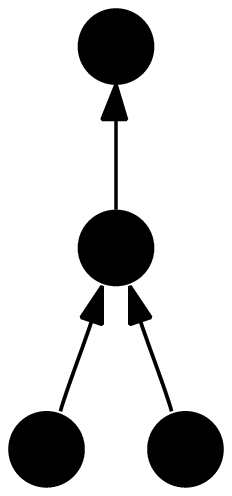} &
  \includegraphics[scale=0.20, angle=180]{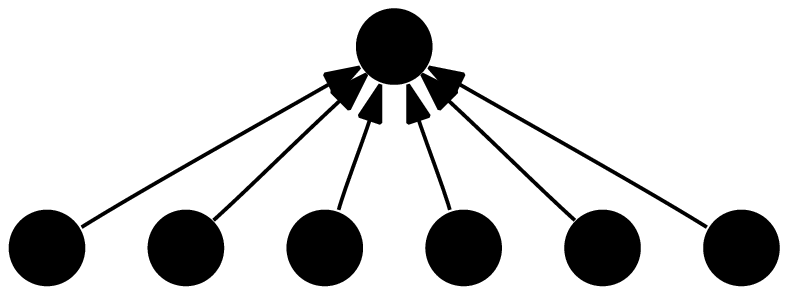} &
  \includegraphics[scale=0.20, angle=180]{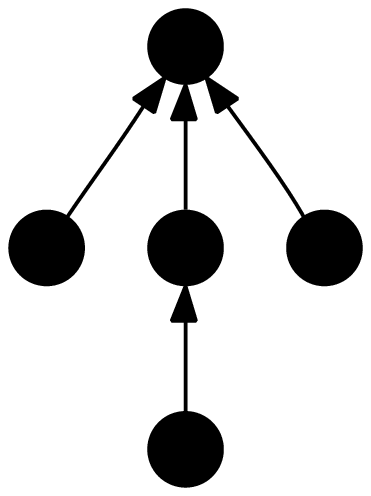} &
  \includegraphics[scale=0.20, angle=180]{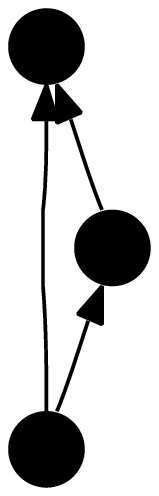} \\
  $G_{2}$ & $G_{3}$ & $G_{4}$ & $G_{5}$ & $G_{6}$ & $G_{7}$ & $G_{8}$ &
    $G_{9}$ & $G_{10}$ & $G_{11}$ & $G_{12}$ \\
  \end{tabular}
  \begin{tabular}{ccccccccccc}
  \includegraphics[scale=0.20, angle=180]{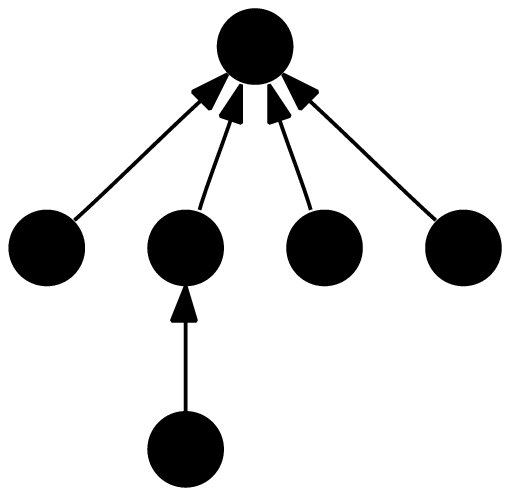} &
  \includegraphics[scale=0.20, angle=180]{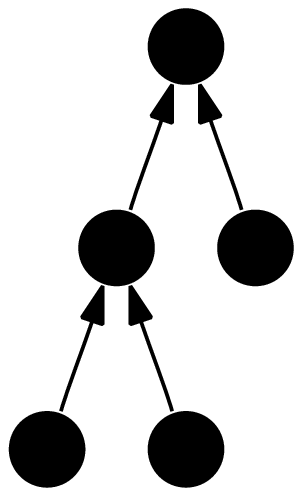} &
  \includegraphics[scale=0.20, angle=180]{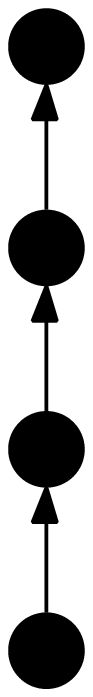} &
  \includegraphics[scale=0.20, angle=180]{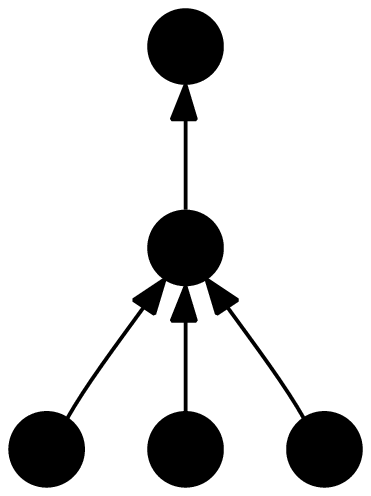} &
  \includegraphics[scale=0.20, angle=180]{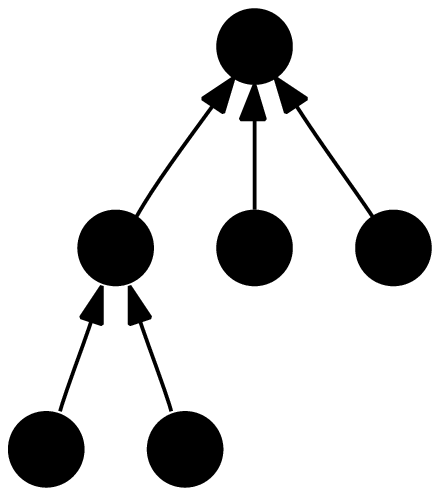} &
  \includegraphics[scale=0.20, angle=180]{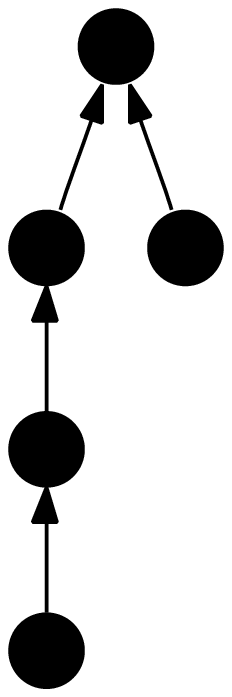} &
  \includegraphics[scale=0.20, angle=180]{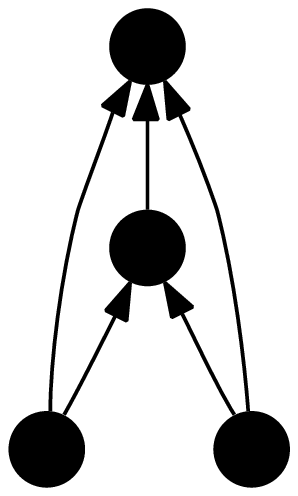} &
  \includegraphics[scale=0.20, angle=180]{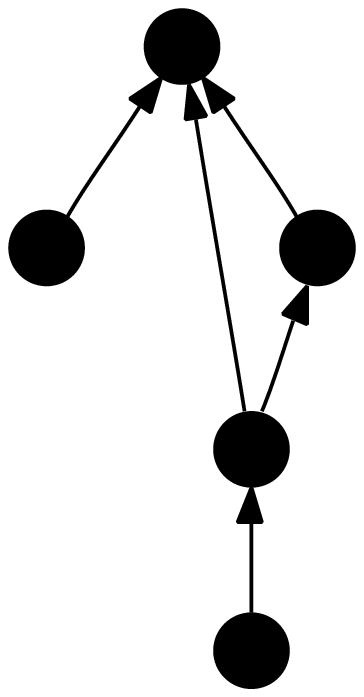} &
  \includegraphics[scale=0.20, angle=180]{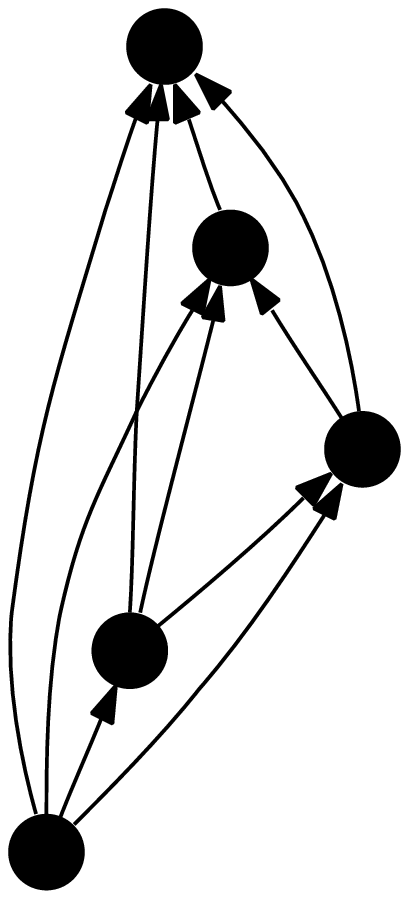} &
  \includegraphics[scale=0.20, angle=180]{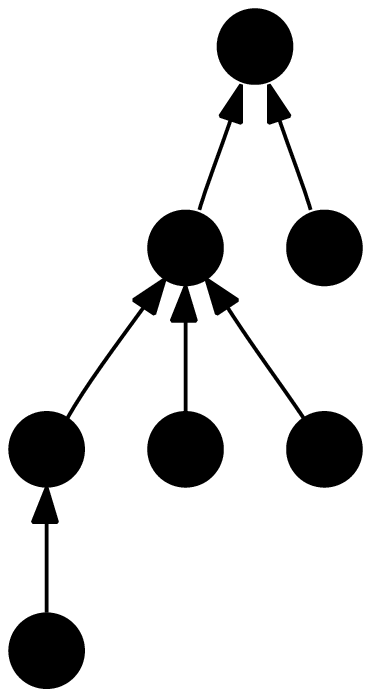} &
  \includegraphics[scale=0.20, angle=180]{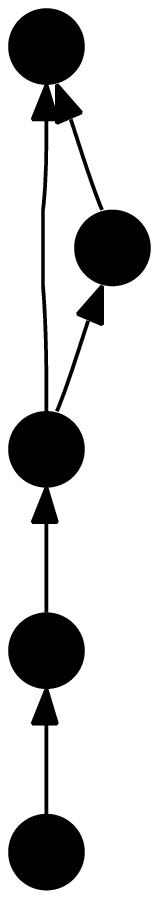} \\
  $G_{14}$ & $G_{15}$ & $G_{16}$ & $G_{18}$ & $G_{29}$ & $G_{34}$ &
  $G_{83}$ & $G_{100}$ & $G_{107}$ & $G_{117}$ & $G_{124}$ \\
  \end{tabular}
  \caption{Common cascade shapes ordered by the frequency.
  Cascade with label $G_r$ has the frequency rank $r$.}
  \label{fig:shapes}
\end{figure*}

\subsection{\PostNet\ topology}

In contrast to \BlogNet\ the \PostNet\ is very sparsely connected. It
contains 2.2 million nodes and only $205,000$ edges. $98\%$ of the posts
are isolated, and the largest connected component accounts for $106,000$
nodes, while the second largest has only 153 nodes.
Figure~\ref{fig:postDeg} shows the in- and out-degree distributions of
the \PostNet\ which follow a power law with exponents $-2.1$ and $-2.9$,
respectively.

\subsection{Patterns in the cascades}
\label{sec:cascades}

We continue with the analysis of the topological aspects of the
information cascades formed when certain posts become popular and are
linked by the other posts. We are especially interested in how this
process propagates, how large are the cascades it forms, and as it will
be shown later, what are the models that mimic cascading behavior and
produce realistic cascades.

Cascades are subgraphs of the \PostNet\ that have a single root post,
are time increasing (source links an existing post), and present the
propagation of the information from the root to the rest of the cascade.

Given the \PostNet\ we extracted all information cascades using the
following procedure. We found all cascade initiator nodes, i.e. nodes
that have zero out-degree, and started following their in-links. This
process gives us a directed acyclic graph with a single root node. As
illustrated in Figure~\ref{fig:cascade} it can happen that two cascades
merge, e.g.\ a post gives a summary of multiple conversations
(cascades). For cascades that overlap our cascade extraction procedure
will extract the nodes bellow the connector node multiple times (since
they belong to multiple cascades). To obtain the examples of the common
shapes and count their frequency we used the algorithms as described
in~\cite{jurij05patterns}.

\subsubsection{Common cascade shapes}

First, we give examples of common \PostNet\ cascade shapes in
Figure~\ref{fig:shapes}. A node represents a post and the influence
flows from the top to the bottom. The top post was written first, other
posts linking to it, and the process propagates. Graphs are ordered by
frequency and the subscript of the label gives frequency rank. Thus,
$G_{124}$ is $124^{th}$ most frequent cascade with 11 occurrences.

We find the total of $2,092,418$ cascades, and 97\% of them are trivial
cascades (isolated posts), 1.8\% are smallest non-trivial cascades
($G_2$), and the remaining 1.2\% of the cascades are topologically more
complex.

Most cascades can essentially be constructed from instances of stars and
trees, which can model more complicated behavior like that shown in
Figure~\ref{fig:shapes}. Cascades tend to be wide, and not too deep.
Structure $G_{107}$, which we call a \emph{cite-all chain}, is
especially interesting. Each post in a chain refers to every post before
it in the chain.

We also find that the cascades found in the graph tend to take certain
shapes preferentially.  Also notice that cascade frequency rank does not
simply decrease as a function of the cascade size. For example, as shown
on Figure~\ref{fig:shapes}, a 4-star ($G_4$) is more common than a chain
of 3 nodes ($G_5$). In general stars and shallow bursty cascades are the
most common type of cascades.

\subsubsection{Cascade topological properties}

What is the common topological pattern in the cascades? We next examine
the general cascade behavior by measuring and characterizing the
properties of real cascades.

\begin{figure*}[t]
  \begin{center}
  \begin{tabular}{c c c}
    \epsfig{file=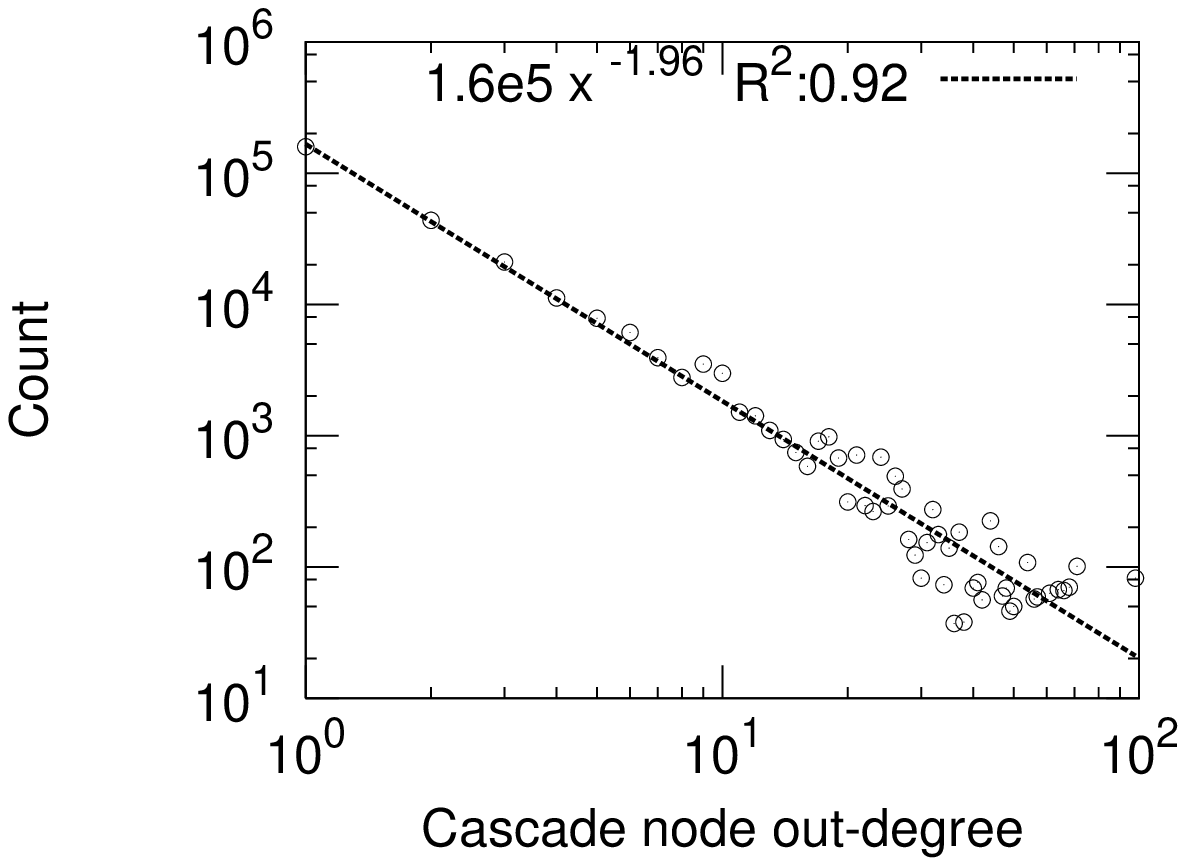, width=0.335\textwidth} &
    \epsfig{file=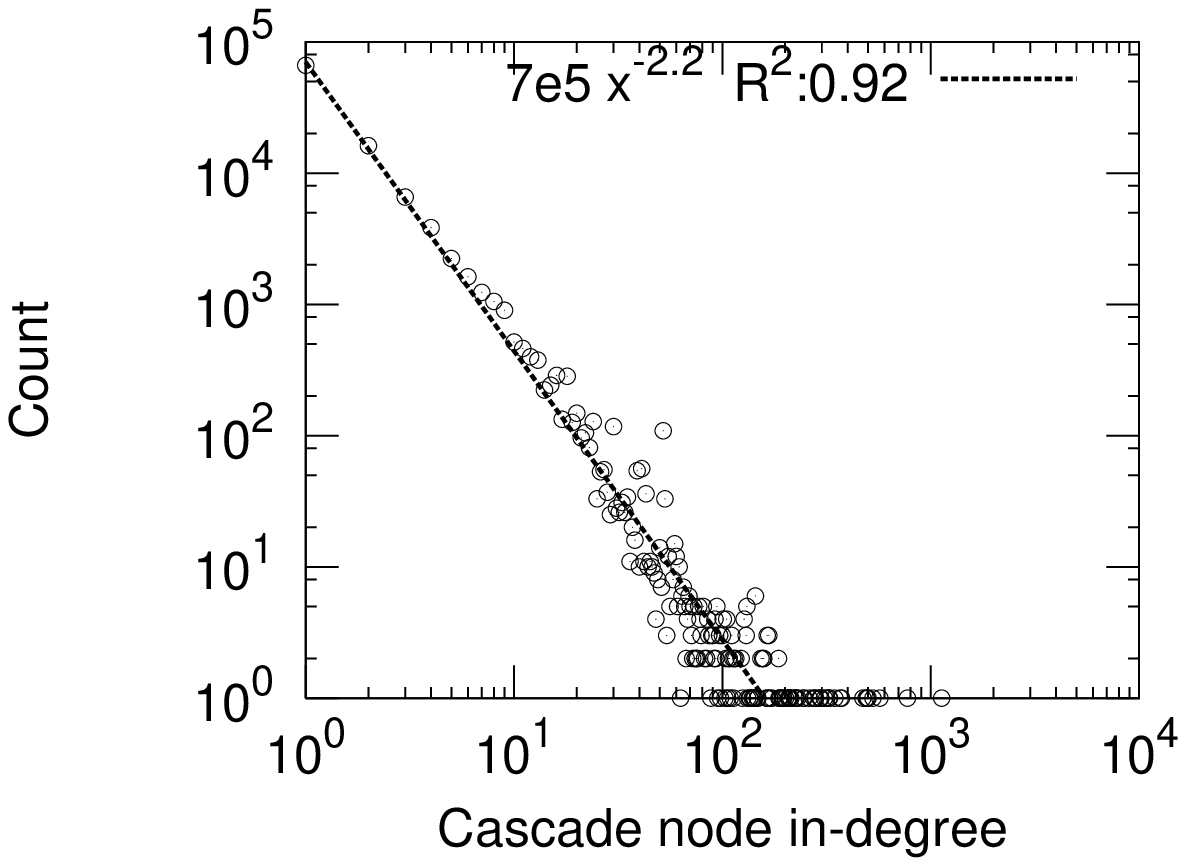, width=0.335\textwidth} &
    \epsfig{file=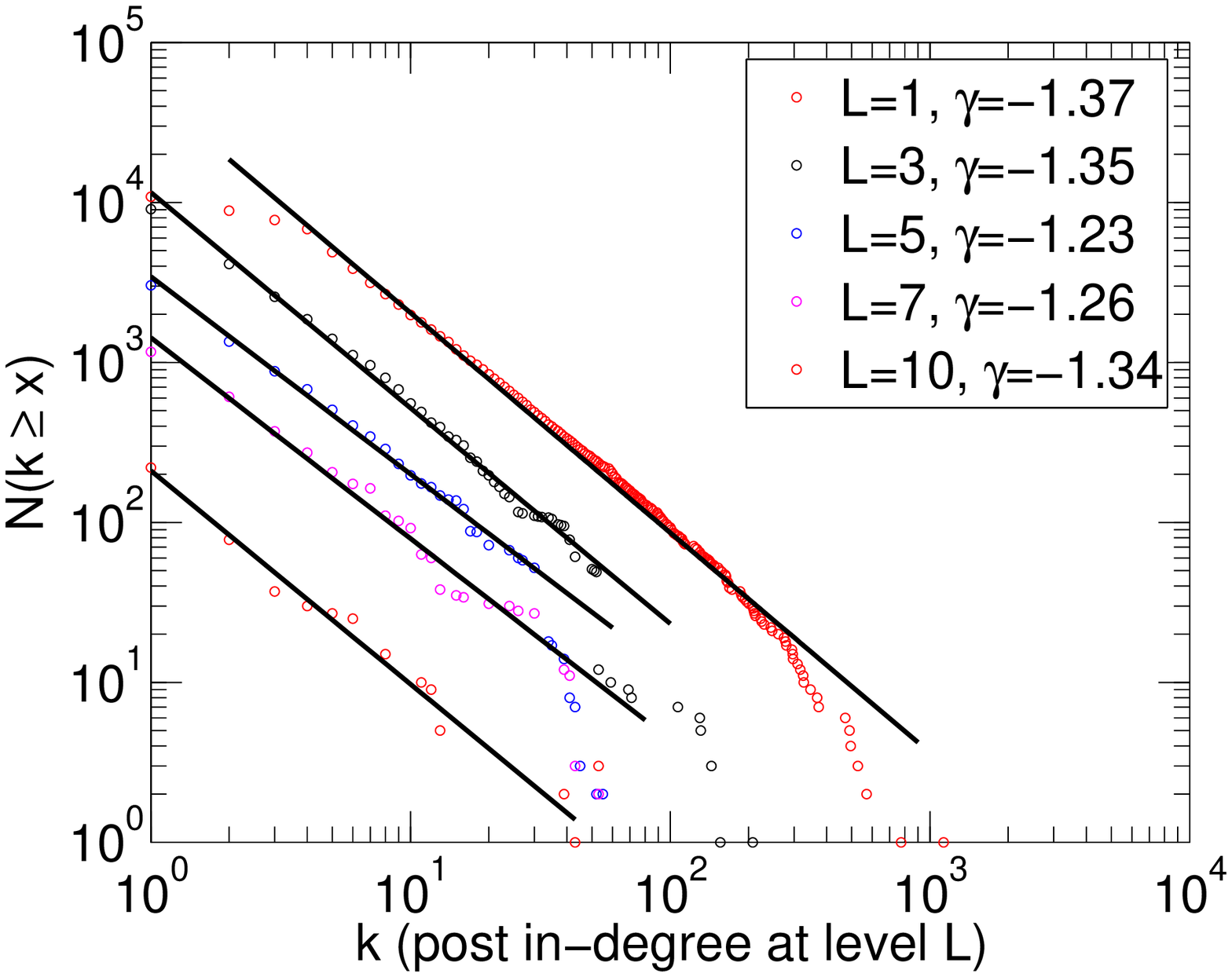, width=0.28\textwidth} \\
    (a) Out-degree & (b) In-degree & (c) In-degree at level $L$\\
  \end{tabular}
  \end{center}
  \caption{Out- and in-degree distribution over all cascades
  extracted from the \PostNet\ (a,b) and the in-degree distribution
  at level $L$ of the cascade. Note all distributions are
  heavy tailed and the in-degree distribution is remarkably stable over
  the levels.}
  \label{fig:cascInOutDeg}
\end{figure*}

First we observe the degree distributions of the cascades. This means
that from the \PostNet\ we extract all the cascades and measure the
overall degree distribution. Essentially we work with a {\em bag of
cascades}, where we treat a cascade as separate disconnected sub-graph
in a large network.

Figure~\ref{fig:cascInOutDeg}(a) plots the out-degree distribution of
the bag of cascades. Notice the cascade out-degree distribution is
truncated, which is the result of not perfect link extraction algorithm
and the upper bound on the post out-degree (500).

Figure~\ref{fig:cascInOutDeg}(b) shows the in-degree distribution of the
bag of cascades, and (c) plots the in-degree distribution of nodes at
level $L$ of the cascade. A node is at level $L$ if it is $L$ hops away
from the root (cascade initiator) node. Notice that the in-degree
exponent is stable and does not change much given the level in the
cascade. This means that posts still attract attention (get linked) even
if they are somewhat late in the cascade and appear towards the bottom
of it.

\begin{figure*}[t]
  \begin{center}
  \begin{tabular}{c c c}
    \epsfig{file=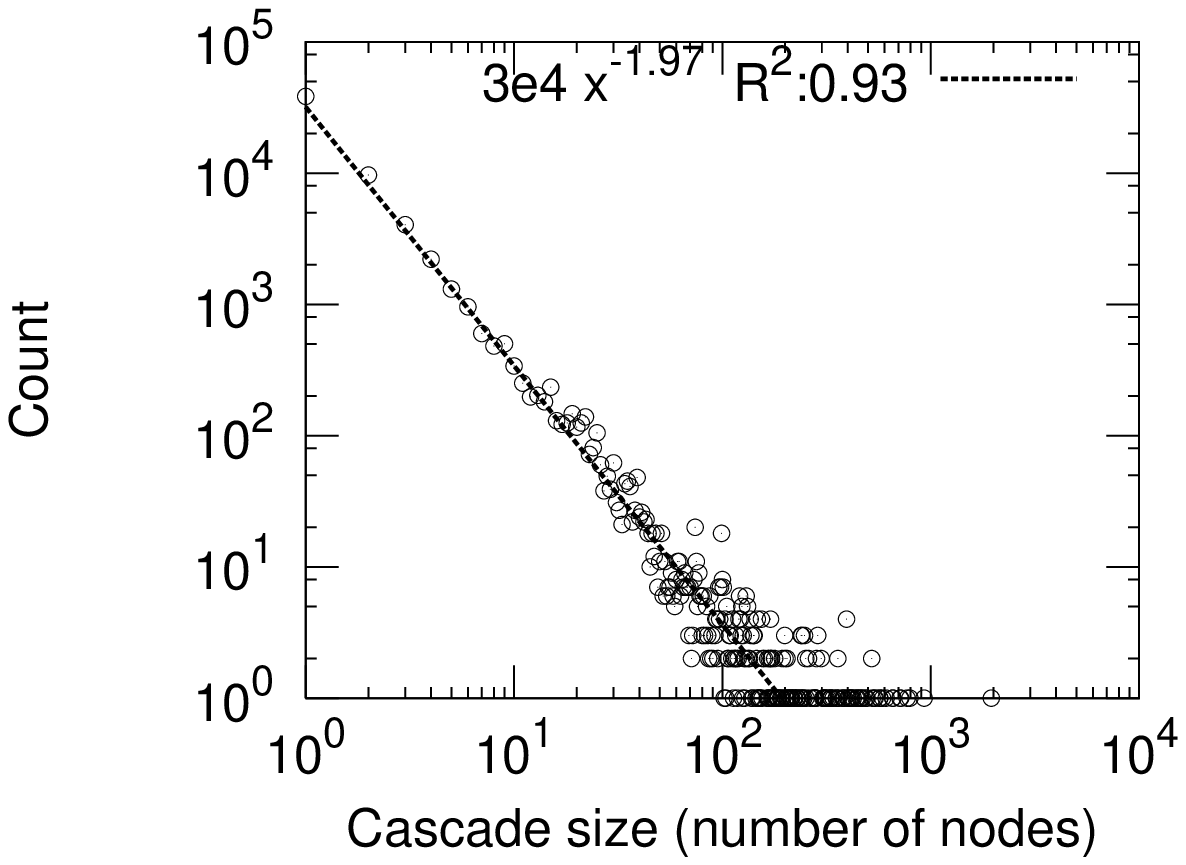, width=0.33\textwidth} &
    \epsfig{file=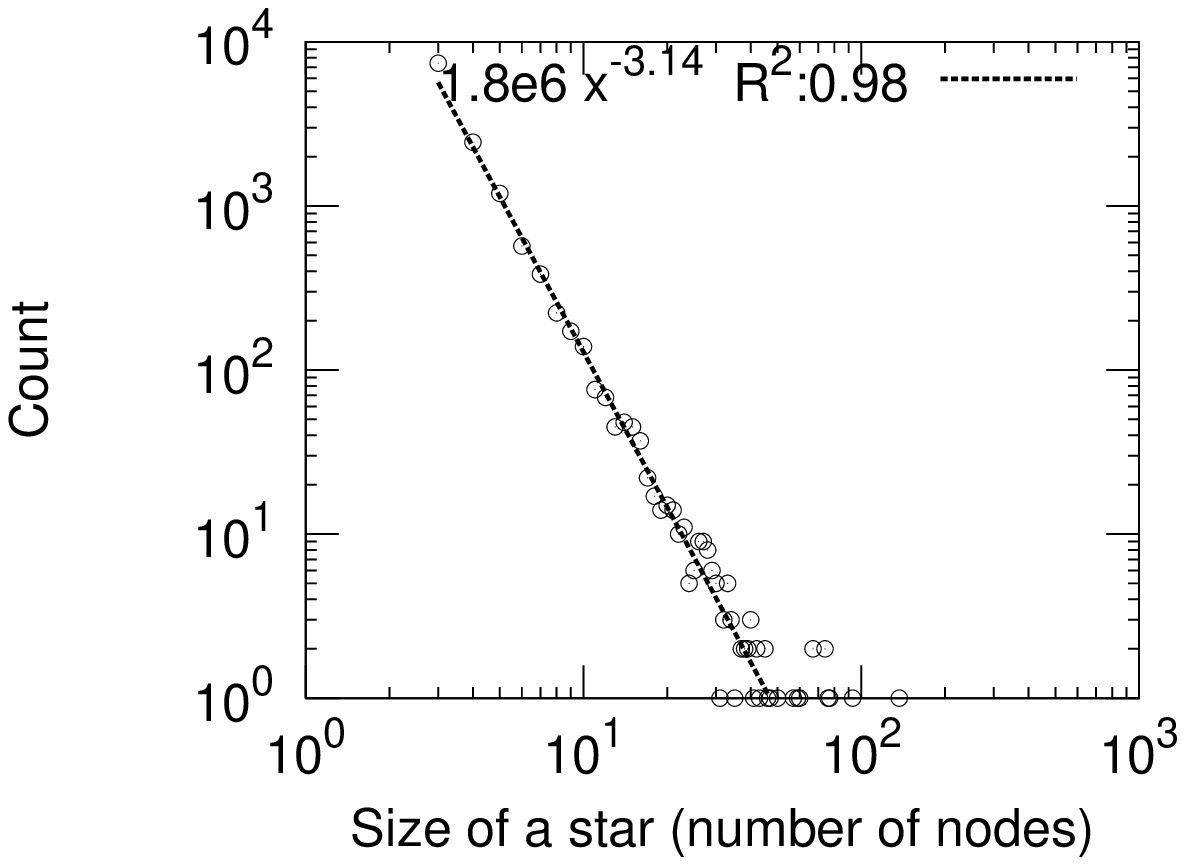, width=0.33\textwidth} &
    \epsfig{file=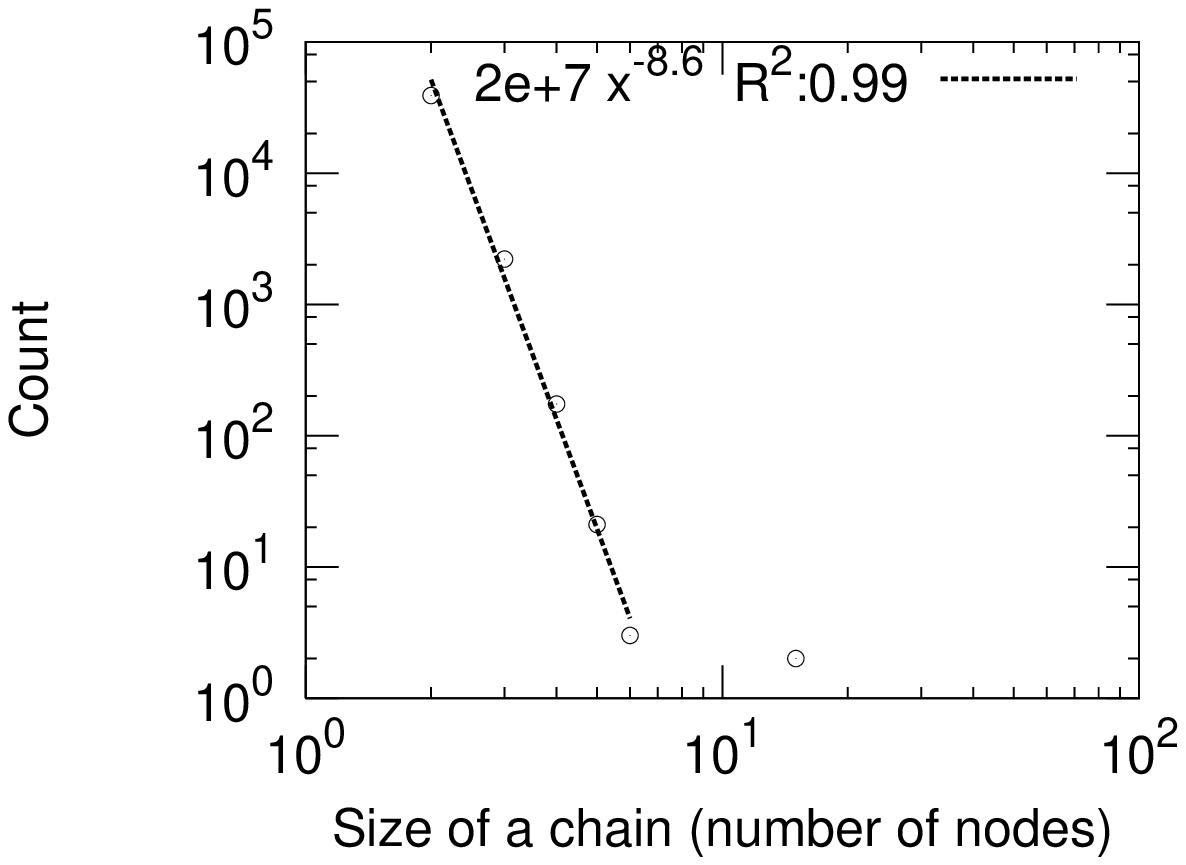, width=0.33\textwidth} \\
    (a) All cascades & (b)  Star cascade & (c) Chain cascade \\
  \end{tabular}
  \end{center}
  \caption{Size distribution over all cascades (a), only stars (b), and chains
  (c). They all follow heavy tailed distributions with increasingly
  steeper slopes.
  }
  \label{fig:cascSzDist}
\end{figure*}

Next, we ask what distribution do cascade sizes follow? Does the
probability of observing a cascade on $n$ nodes decreases exponentially
with $n$? We examine the {\em Cascade Size Distributions} over the bag
of cascades extracted from the \PostNet. We consider three different
distributions: over all cascade size distribution, and separate size
distributions of star and chain cascades. We chose stars and chains
since they are well defined, and
given the number of nodes in the cascade, there is no ambiguity in the
topology of a star or a chain.

Figure~\ref{fig:cascSzDist} gives the Cascade Size Distribution plots.
Notice all follow a heavy-tailed distribution. We fit a power-law
distribution and observe that overall cascade size distribution has
power-law exponent of $\approx -2$ (Figure~\ref{fig:cascSzDist}(a)),
stars have $\approx -3.1$ (Figure~\ref{fig:cascSzDist}(b)), and chains
are small and rare and decay with exponent $\approx -8.5$
(Fig.~\ref{fig:cascSzDist}(c)). Also notice there are outlier chains
(Fig.~\ref{fig:cascSzDist}(c)) that are longer than expected. We
attribute this to possible flame wars between the blogs, where authors
publish posts and always refer to the last post of the other author.
This creates chains longer than expected.

\begin{observation}
Probability of observing a cascade on $n$ nodes follows a Zipf
distribution:
\[
  p(n) \propto n^{-2}
\]
\end{observation}

\begin{figure}[t]
  \begin{center}
  \begin{tabular}{c c}
    \epsfig{file=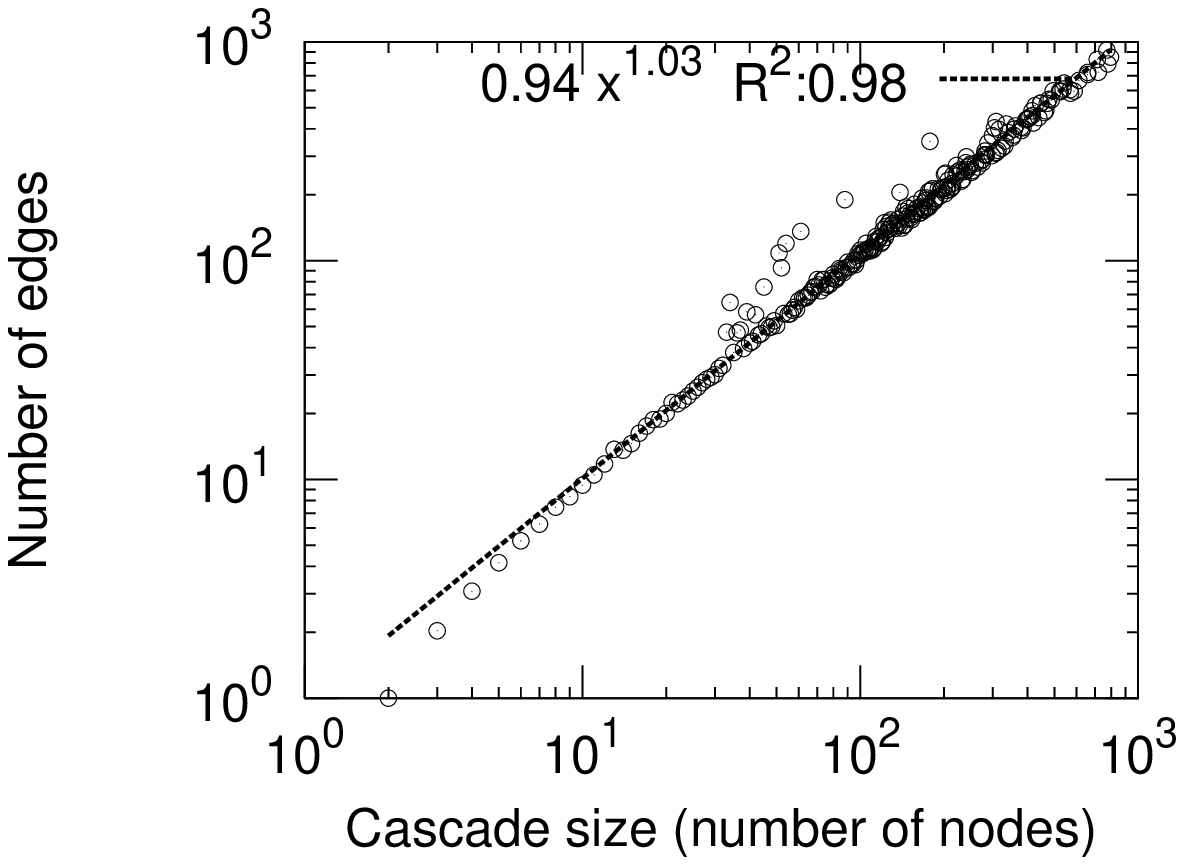, width=0.45\textwidth} &
    \epsfig{file=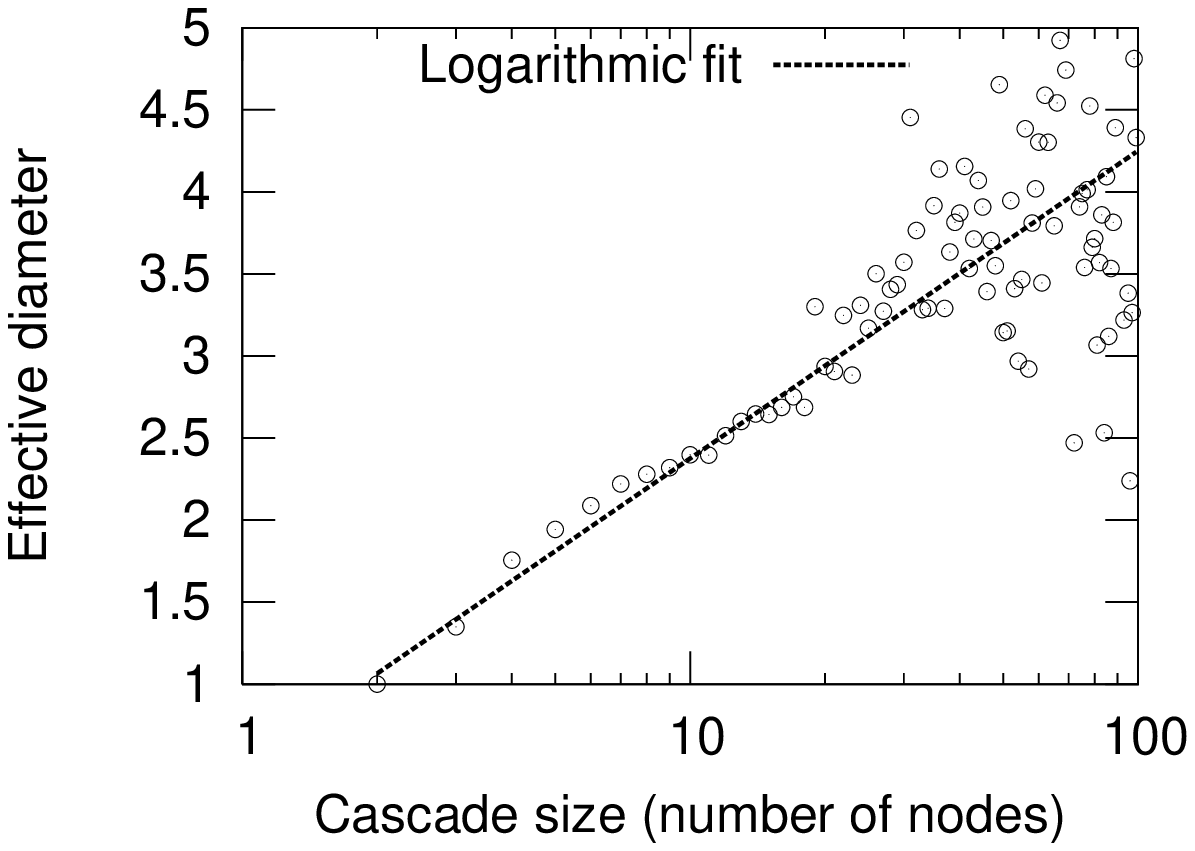, width=0.45\textwidth} \\
    (a)  Edges vs. nodes & (b) Diameter \\
  \end{tabular}
  \end{center}
  \caption{Diameter and the number of edges vs. the cascade size.
  Notice that diameter increases logarithmically with the cascade
  size, while the number of edges basically grows linearly with the
  cascade size. This suggests cascades are mostly tree-like structures.}
  \label{fig:cascDplDiam}
\end{figure}

As suggested by Figure~\ref{fig:shapes} most cascades follow tree-like
shapes. To further verify this we examine how the diameter, defined as
the length of the longest undirected path in the cascade, and the
relation between the number of nodes and the number of edges in the
cascade change with the cascade size in Figure~\ref{fig:cascDplDiam}.

This gives further evidence that the cascades are mostly tree-like. We
plot the number of nodes in the cascade vs.\ the number of edges in the
cascade in Figure~\ref{fig:cascDplDiam}(a). Notice the number of edges
$e$ in the cascade increases almost linearly with the number of nodes
$n$ ($e \propto n^{1.03}$). This suggests that the average degree in the
cascade remains constant as the cascade grows, which is a property of
trees and stars. Next, we also measure cascade diameter vs.\ cascade
size (Figure~\ref{fig:cascDplDiam}(b)). We plot on linear-log scales and
fit a logarithmic function. Notice the diameter increases
logarithmically with the size of the cascade, which means the cascade
needs to grow exponentially to gain linear increase in diameter, which
is again a property of the balanced trees and very sparse graphs.

\subsubsection{Collisions of cascades}

By the definition we adopt in this paper, the cascade has a single
initiator node, but in real life one would also expect that cascades
collide and merge. There are connector nodes which are the first to
bring together separate cascades. As the cascades merge, all the nodes
bellow the connector node now belong to multiple cascades. We measure
the distribution over the connector nodes and the nodes that belong to
multiple cascades.

\begin{figure}[h]
  \begin{center}
  \begin{tabular}{c c}
    \epsfig{file=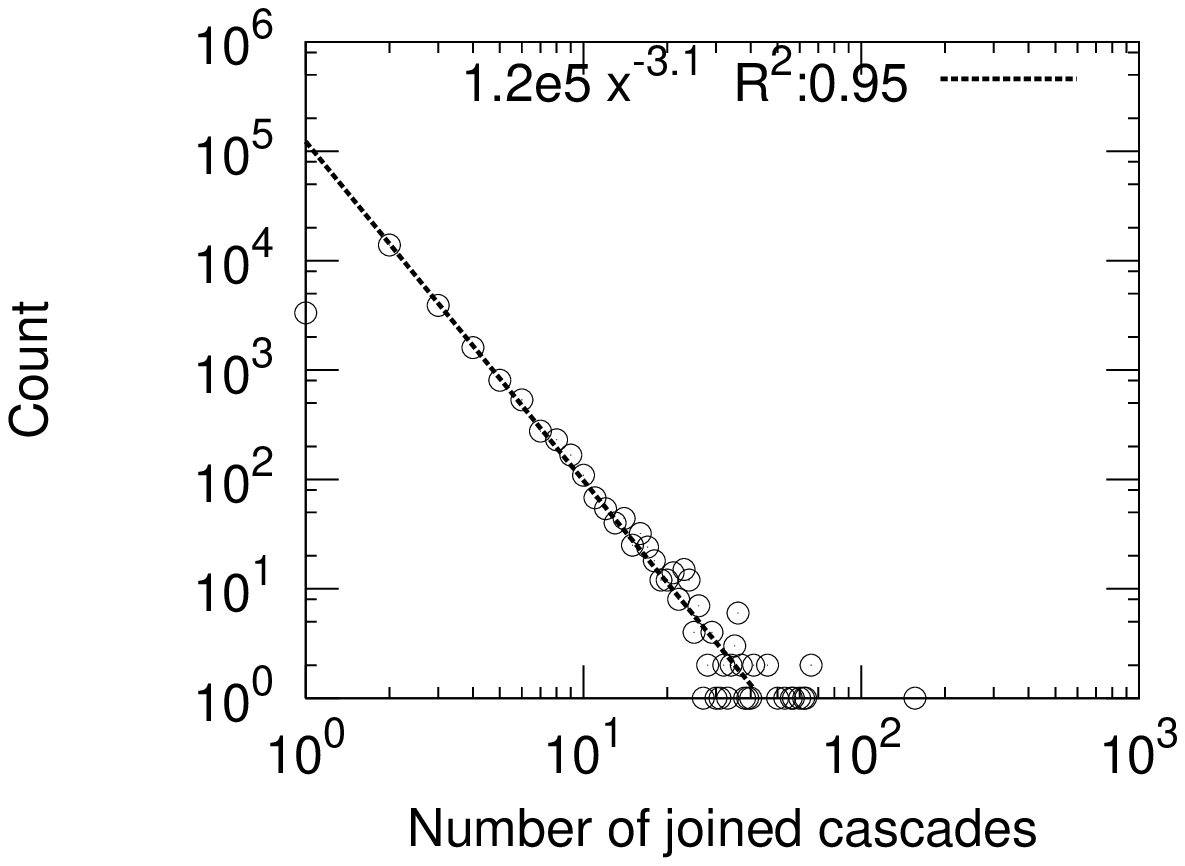, width=0.45\textwidth} &
    \epsfig{file=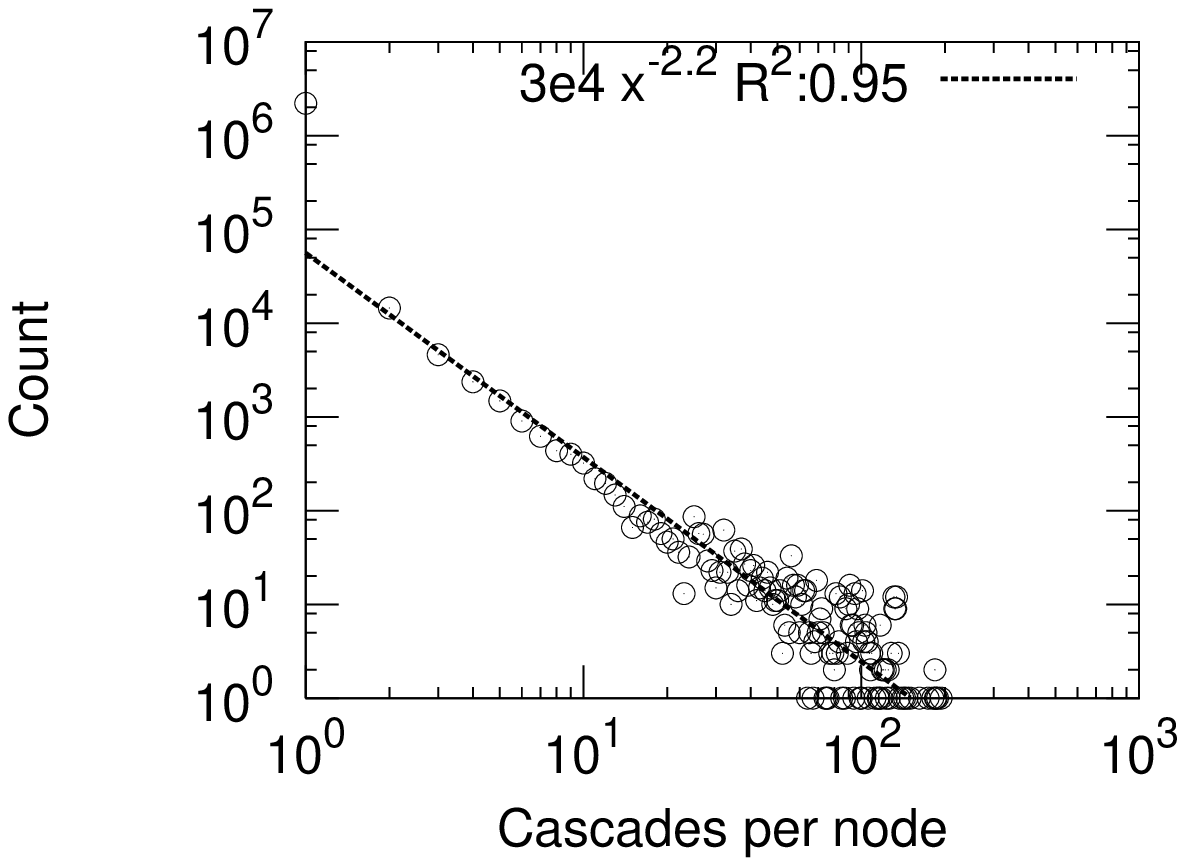, width=0.45\textwidth} \\
  \end{tabular}
  \end{center}
  \caption{Distribution of joined cascades by the connector nodes
  (a). We only consider nodes with out-degree greater than 1.
  Distribution of a number of cascades a post belongs to (b);
  $98\%$ of posts belong to a single cascade.}
  \label{fig:connectors}
\end{figure}

First, we consider only the connector nodes and plot the distribution
over how many cascades a connector joins
(Figure~\ref{fig:connectors}(a)). We only consider nodes with out-degree
greater than 1, since nodes with out-degree 1 are trivial connectors --
they are connecting the cascade they belong to. But there are still
posts that have out-degree greater than 1, and connect only one cascade.
These are the posts that point multiple out-links inside the same
cascade (e.g. $G_{12}$ and $G_{107}$ of Figure~\ref{fig:shapes}). The
dip the at the number of joined cascades equal to 1 in
Figure~\ref{fig:connectors}(a) gives the number of such nodes.

As cascades merge, all the nodes that follow belong to multiple
cascades. Figure~\ref{fig:connectors}(b) gives the distribution over the
number of cascades a node belongs to. Here we consider all the nodes and
find out that $98\%$ of all nodes belong to a single cascade, and the
rest of distribution follows a power-law with exponent $-2.2$.

\section{Proposed model and insights}
\label{sec:models}

What is the underlying hidden process that generates cascades? Our goal
here is to find a generative model that generates cascades with
properties observed in section~\ref{sec:cascades}
(Figures~\ref{fig:cascInOutDeg} and~\ref{fig:cascSzDist}). We aim for
simple and intuitive model with the least possible number of parameters.

\subsection{\CGM}

We present a conceptual model for generating information cascades that
produces cascade graphs matching several properties of real cascades.
Our model is intuitive and requires only a single parameter that
corresponds to how interesting (easy spreading) are the conversations in
general on the blogosphere.

Intuitively, cascades are generated by the following principle. A post
is posted at some blog, other bloggers read the post, some create new
posts, and link the source post. This process continues and creates a
cascade. One can think of cascades being a graph created by the spread
of the virus over the \BlogNet. This means that the initial post
corresponds to infecting a blog. As the cascade unveils, the virus
(information) spreads over the network and leaves a trail. To model this
process we use a single parameter $\B$ that measures how infectious are
the posts on the blogosphere. Our model is very similar to the SIS
(susceptible -- infected -- susceptible) model from the
epidemiology~\cite{Hethcote:2000}.

Next, we describe the model. Each blog is in one of two states: {\em
infected} or {\em susceptible}. If a blog is in the infected state this
means that the blogger just posted a post, and the blog now has a chance
to spread its influence. Only blogs in the susceptible (not infected)
state can get infected. When a blog successfully infects another blog, a
new node is added to the cascade, and an edge is created between the
node and the source of infection. The source immediately recovers, i.e.
a node remains in the infected state only for one time step. This gives
the model ability to infect a blog multiple times, which corresponds to
multiple posts from the blog participating in the same cascade.

More precisely, a single cascade of the {\em \CGM} is generated by the
following process.
\begin{enumerate}
  \item[(i)] Uniformly at random pick blog $u$ in the \BlogNet\ as a
      starting point of the cascade, set its state to {\em
      infected}, and add a new node $u$ to the cascade graph.
  \item[(ii)] Blog $u$ that is now in infected state, infects each
      of its uninfected directed neighbors in the \BlogNet\
      independently with probability $\B$. Let $\{v_1,\dots, v_n\}$
      denote the set of infected neighbors.
  \item[(iii)] Add new nodes $\{v_1,\dots, v_n\}$ to the cascade and
      link them to node $u$ in the cascade.
  \item[(iv)] Set state of node $u$ to not infected. Continue
      recursively with step (ii) until no nodes are infected.
\end{enumerate}

We make a few observations about the proposed model. First, note that
the blog immediately recovers and thus can get infected multiple times.
Every time a blog gets infected a new node is added to the cascade. This
accounts for multiple posts from the blog participating in the same
cascade. Second, we note that in this version of the model we do not try
to account for topics or model the influence of particular blogs. We
assume that all blogs and all conversations have the same value of the
parameter $\B$. Third, the process as describe above generates cascades
that are trees. This is not big limitation since we observed that most
of the cascades are trees or tree-like. In the spirit of our notion of
cascade we assume that cascades have a single starting point, and do not
model for the collisions of the cascades.

\subsection{Validation of the model}

We validate our model by extensive numerical simulations. We compare the
obtained cascades towards the real cascades extracted from the \PostNet.
We find that the model matches the cascade size and degree
distributions.

We use the real \BlogNet\ over which we propagate the cascades. Using
the \CGM\ we also generate the same number of cascades as we found in
{\PostNet} ($\approx 2$ million). We tried several values of $\B$
parameter, and at the end decided to use $\B = 0.025$. This means that
the probability of cascade spreading from the infected to an uninfected
blog is $2.5\%$. We simulated our model 10 times, each time with a
different random seed, and report the average.

\begin{figure}[t]
  \centering
  \begin{tabular}{ccccccccccc}
  \includegraphics[scale=0.30, angle=180]{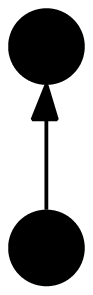} &
  \includegraphics[scale=0.30, angle=180]{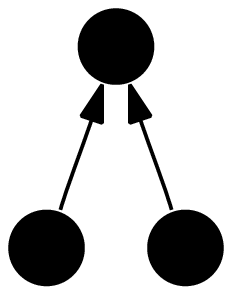} &
  \includegraphics[scale=0.30, angle=180]{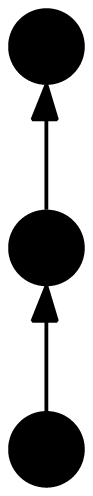} &
  \includegraphics[scale=0.30, angle=180]{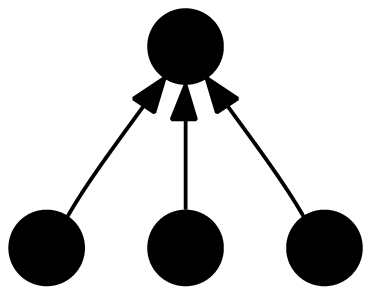} &
  \includegraphics[scale=0.30, angle=180]{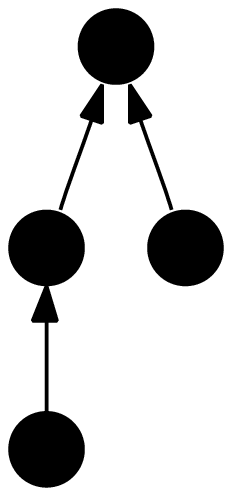} &
  \includegraphics[scale=0.30, angle=180]{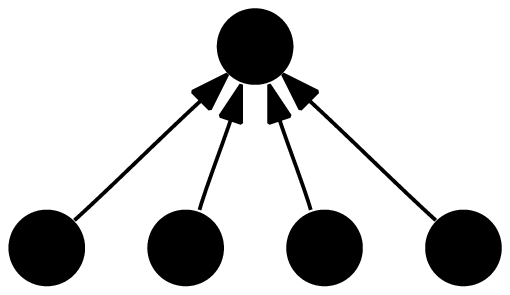} &
  \includegraphics[scale=0.30, angle=180]{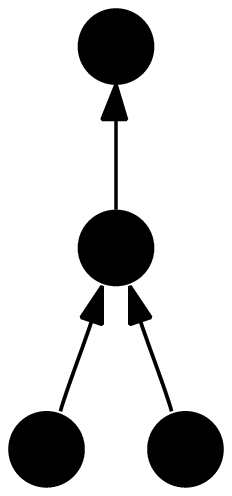} &
  \includegraphics[scale=0.30, angle=180]{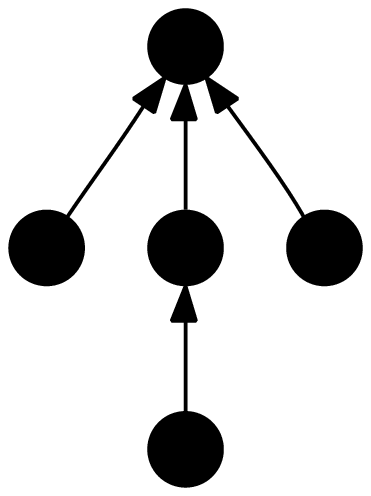} &
  \includegraphics[scale=0.30, angle=180]{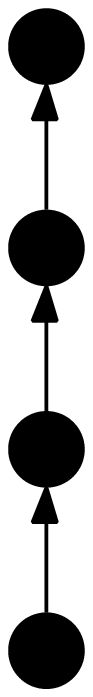} \\
  \end{tabular}
  \caption{Top 10 most frequent cascades as generated by the \CGM.
  Notice similar shapes and frequency ranks as in Figure~\ref{fig:shapes}.}
  \label{fig:cgmShapes}
\end{figure}

First, we show the top 10 most frequent cascades (ordered by frequency
rank) as generated by the \CGM\ in Figure~\ref{fig:cgmShapes}. Comparing
them to most frequent cascades from Figure~\ref{fig:shapes} we notice
that top 7 cascades are matched exactly (with an exception of ranks of
$G_4$ and $G_5$ swapped), and rest of cascades can also be found in real
data.

\begin{figure*}[thb]
  \begin{center}
  \begin{tabular}{c c}
    \epsfig{file=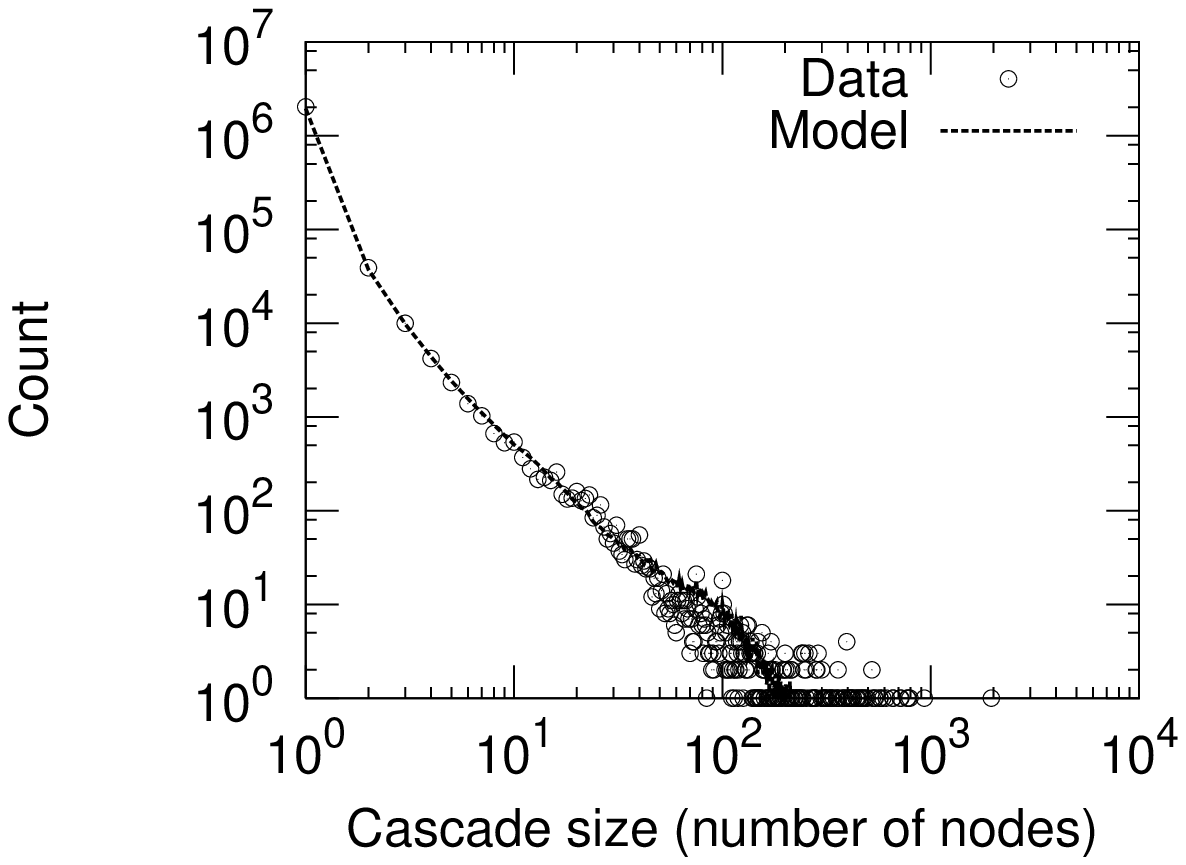, width=0.45\textwidth} &
    \epsfig{file=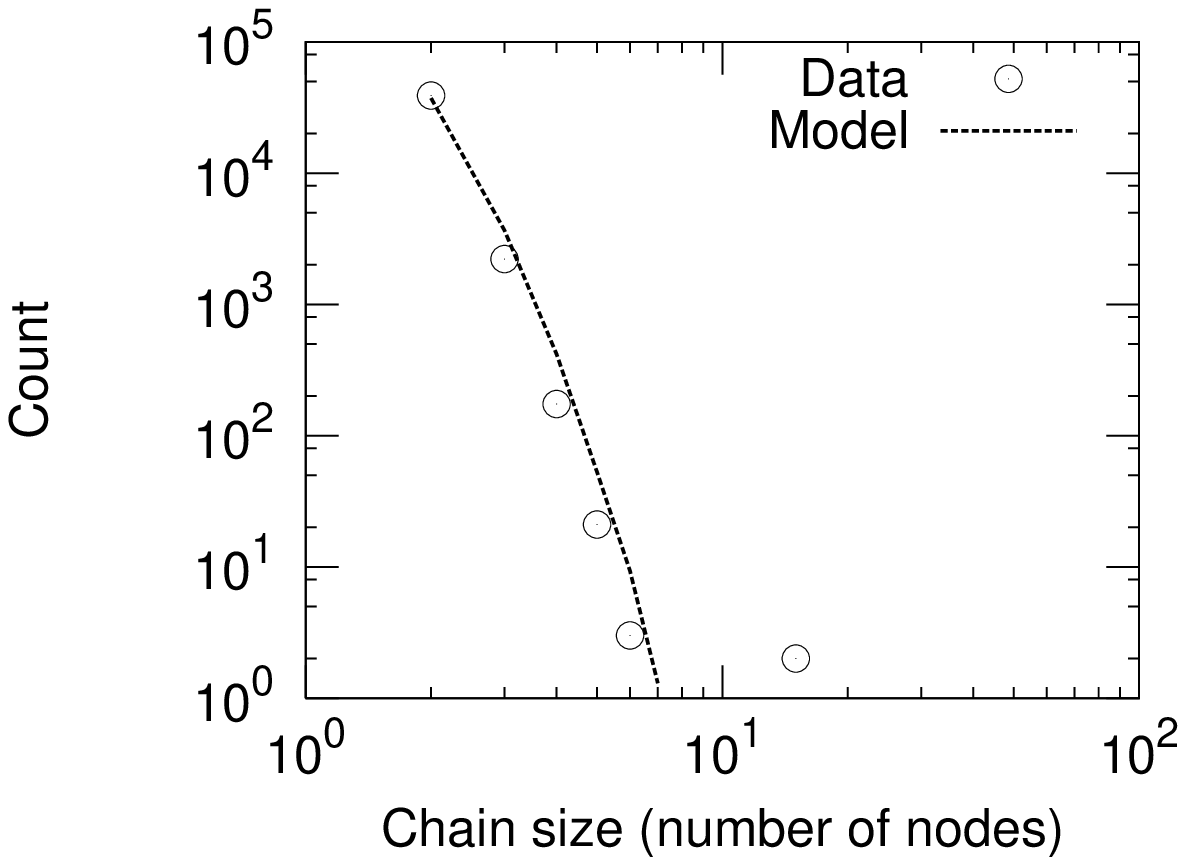, width=0.45\textwidth} \\
    (a) All cascades & (b) Chain cascades \\
    \epsfig{file=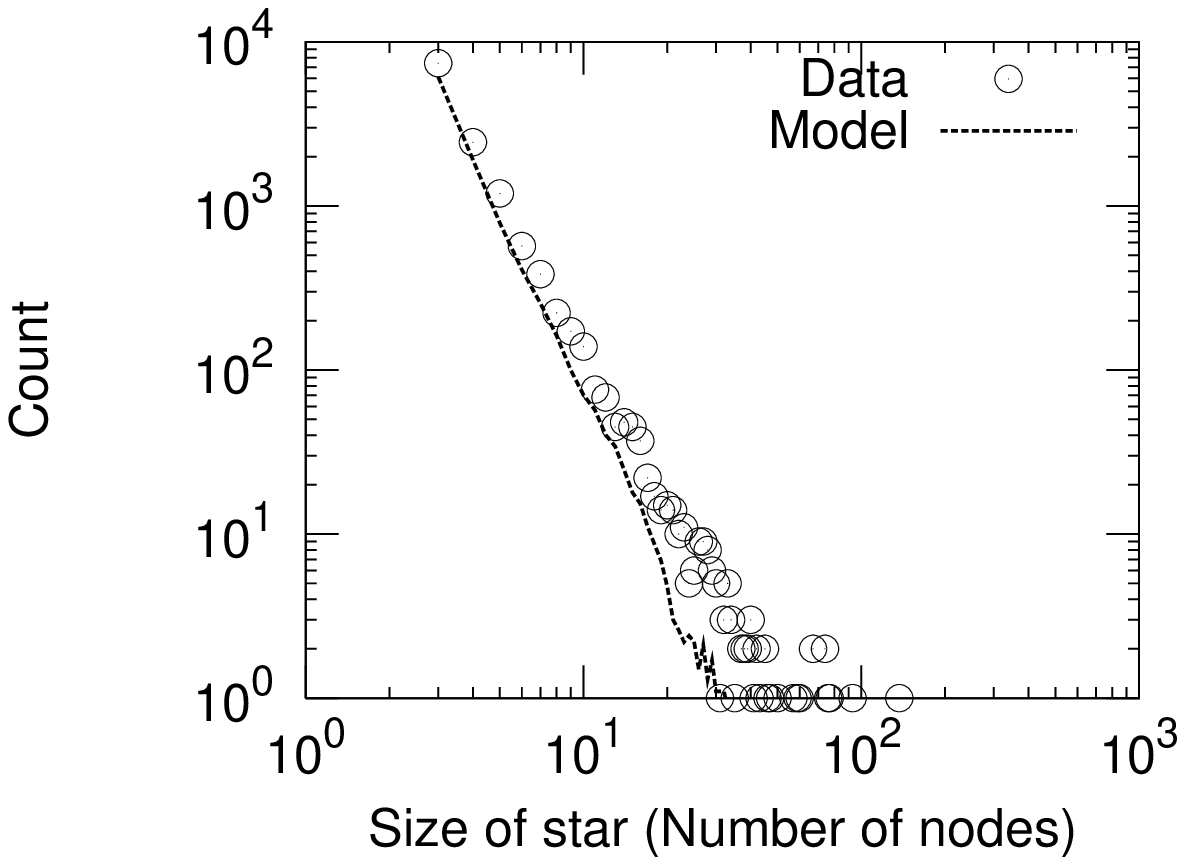, width=0.45\textwidth} &
    \epsfig{file=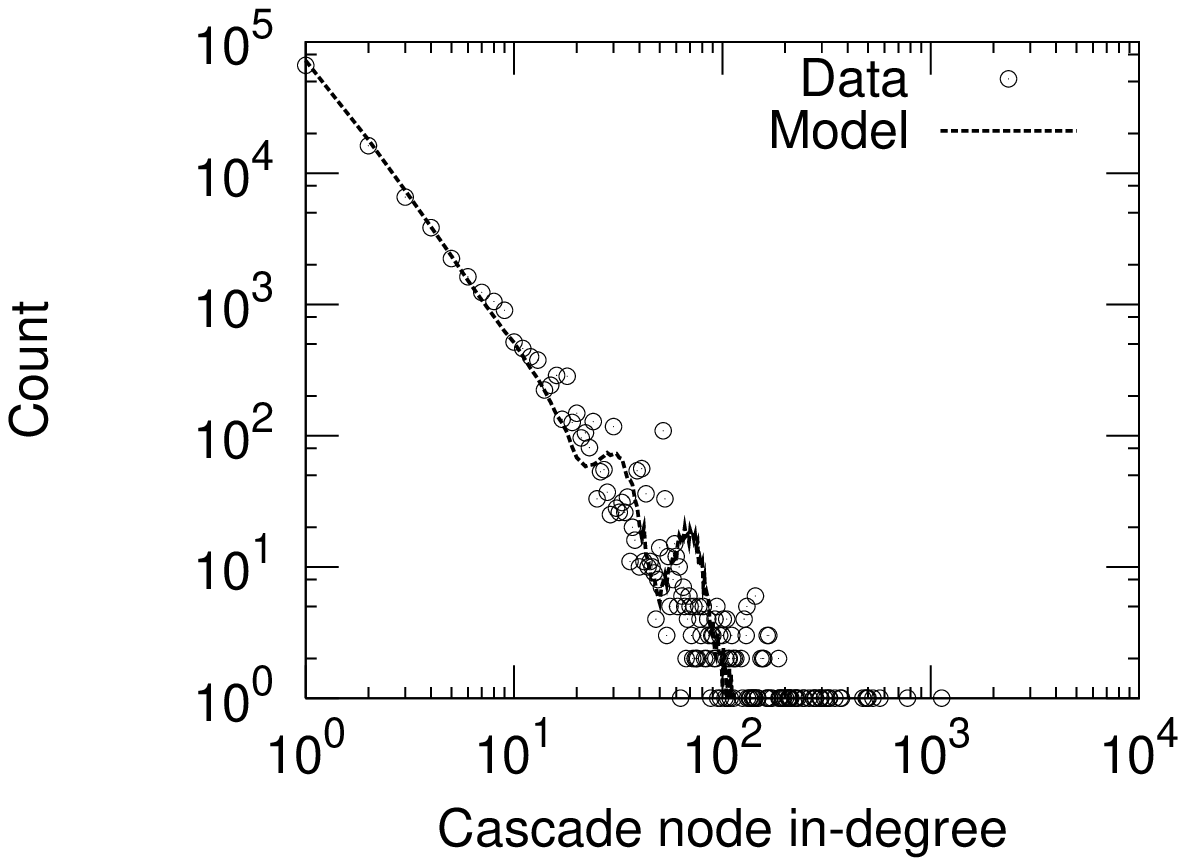, width=0.45\textwidth} \\
    (c)  Star cascades & (d) In-degree distribution\\
  \end{tabular}
  \end{center}
  \caption{Comparison of the true data and the model. We plotted the
  distribution of the true cascades with circles and the estimate of
  our model with dashed line. Notice remarkable agreement between the
  data and the prediction of our simple model.}
  \label{fig:modelRes}
\end{figure*}

Next, we show the results on matching the cascade size and degree
distributions in Figure~\ref{fig:modelRes}. We plot the true
distributions of the cascades extracted from the \PostNet\ with dots,
and the results of our model are plotted with a dashed line. We compare
four properties of cascades: (a) overall cascade size distribution, (b)
size distribution of chain cascades, (c) size distribution of stars, and
(d) in-degree distribution over all cascades.

Notice a very good agreement between the reality and simulated cascades
in all plots. The distribution over of cascade sizes is matched best.
Chains and stars are slightly under-represented, especially in the tail
of the distribution where the variance is high. The in-degree
distribution is also matched nicely, with an exception of a spike that
can be attributed to a set of outlier blogs all with in-degree 52. Note
that cascades generated by the \CGM\ are all trees, and thus the
out-degree for every node is 1.

\subsection{Variations of the model}

We also experimented with other, more sophisticated versions of the
model. Namely, we investigated various strategies of selecting a
starting point of the cascade, and using edge weights (number of
blog-to-blog links) to further boost cascades.

We considered selecting a cascade starting blog based on the blog
in-degree, in-weight or the number of posts. We experimented variants
where the probability $\B$ of propagating via a link is not constant but
also depends on the weight of the link (number of references between the
blogs). We also considered versions of the model where the probability
$\B$ exponentially decays as the cascade spreads away from the origin.

We found out that choosing a cascade starting blog in a biased way
results in too large cascades and non-heavy tailed distributions of
cascade sizes. Intuitively, this can be explained by the fact that
popular blogs are in the core of the \BlogNet, and it is very easy to
create large cascades when starting in the core. A similar problem
arises when scaling $\B$ with the edge weight. This can also be
explained by the fact that we are not considering specific topics and
associate each edge with a topic (some blog-to-blog edges may be very
topic-specific) and thus we allow the cascade to spread over all edges
regardless of the particular reason (the topic) that the edge between
the blogs exists. This is especially true for blogs like BoingBoing
\footnote{\url{www.boingboing.net}} that are very general and just a
collection of ``wonderful things''.

\section{Discussion}
\label{sec:discussion} Our finding that the the popularity of posts
drops off with a power law distribution is interesting since intuition
might lead one to believe that people would ``forget'' a post topic in
an exponential pattern. However, since linking patterns are based on the
behaviors of individuals over several instances, much like other
real-world patterns that follow power laws such as traffic to Web pages
and scientists' response times to letters~\cite{Vazquez:2006}, it is
reasonable to believe that a high number of individuals link posts
quickly, and later linkers fall off with a heavy-tailed pattern.

Our findings have potential applications in many areas. One could argue
that the conversation mass metric, defined as the total number of posts
in all conversation trees below the point in which the blogger
contributed, summed over all conversation trees in which the blogger
appears, is a better proxy for measuring influence. This metric captures
the mass of the total conversation generated by a blogger, while number
of in-links captures only direct responses to the blogger's posts.

For example, we found that BoingBoing, which a very popular blog about
amusing things, is engaged in many cascades. Actually, 85\% of all
BoingBoing posts were cascade initiators. The cascades generally did not
spread very far but were wide (e.g., $G_{10}$ and $G_{14}$ in
Fig.~\ref{fig:shapes}). On the other hand $53\%$ of posts from a
political blog MichelleMalkin~\footnote{\url{www.michellemalkin.com}}
were cascade initiators. But the cascade here were deeper and generally
larger (e.g., $G_{117}$ in Fig.~\ref{fig:shapes}) than those of
BoingBoing.

\section{Conclusion}
\label{sec:conclusion}

We analyzed one of the largest available collections of blog
information, trying to find how blogs behave and how information
propagates through the blogosphere.  We studied two structures, the
``\BlogNet''  and the ``\PostNet''. Our contributions are two-fold: (a)
The discovery of a wealth of temporal and topological patterns and (b)
the development of a generative model that mimics the behavior of real
cascades. In more detail, our findings are summarized as follows:

\begin{itemize}
\item {\em Temporal Patterns:} The decline of a post's popularity
    follows a power law. The slope is $\approx$-1.5, the slope
    predicted by a very recent theory of heavy tails in human
    behavior~\cite{barabasi05human}
\item {\em Topological Patterns:} Almost any metric we examined
    follows a power law: size of cascades, size of blogs, in- and
    out-degrees. To our surprise, the number of in- and out-links of
    a blog are not correlated. Finally, stars and chains are basic
    components of cascades, with stars being more common.
\item {\em Generative model:} Our idea is  to reverse-engineer the
    underlying social network of blog-owners, and to treat the
    influence propagation between blog-posts as a flu-like virus,
    that is, the SIS model in epidemiology. Despite its simplicity,
    our model generates cascades that match very well the real
    cascades with respect to in-degree distribution, cascade size
    distribution, and popular cascade shapes.
\end{itemize}

Future research could try to include the content of the posts, to help
us find even more accurate patterns of influence propagation. Another
direction is to spot anomalies and link-spam attempts, by noticing
deviations from our patterns.

\section*{Acknowledgements} This material is based upon work
supported by the National Science Foundation under Grants No.
SENSOR-0329549 medium ITR IIS-0534205 and under the auspices of the U.S.
Department of Energy by University of California Lawrence Livermore
National Laboratory under contract No.W-7405-ENG-48. This work is also
partially supported by the Pennsylvania Infrastructure Technology
Alliance (PITA), an IBM Faculty Award, a Yahoo Research Alliance Gift,
with additional funding from Intel and NTT.    Additional funding was
provided by a generous gift from Hewlett-Packard. Jure Leskovec was
partially supported by a Microsoft Research Graduate Fellowship, and
Mary McGlohon by a National Science Foundation Graduate Fellowship.

Any opinions, findings, and conclusions or recommendations expressed in
this material are those of the author(s) and do not necessarily reflect
the views of the National Science Foundation, or other funding parties.


\begin{thebibliography}{10}

\bibitem{Lada05election}
L.~A. Adamic and N.~Glance.
\newblock The political blogosphere and the 2004 u.s. election: divided they
  blog.
\newblock In {\em LinkKDD '05: Proceedings of the 3rd international workshop on
  Link discovery}, pages 36--43, 2005.

\bibitem{Adar:2005}
E.~Adar and L.~A. Adamic.
\newblock Tracking information epidemics in blogspace., 2005.

\bibitem{Bailey1975Diseases}
N.~Bailey.
\newblock {\em The Mathematical Theory of Infectious Diseases and its
  Applications}.
\newblock Griffin, London, 1975.

\bibitem{barabasi05human}
A.-L. Barabasi.
\newblock The origin of bursts and heavy tails in human dynamics.
\newblock {\em Nature}, 435:207, 2005.

\bibitem{Bikchandani:1992}
S.~Bikhchandani, D.~Hirshleifer, and I.~Welch.
\newblock A theory of fads, fashion, custom, and cultural change in
  informational cascades.
\newblock {\em Journal of Political Economy}, 100(5):992--1026, October 1992.

\bibitem{Crovella96Self}
M.~Crovella and A.~Bestavros.
\newblock Self-similarity in world wide web traffic, evidence and possible
  causes.
\newblock {\em Sigmetrics}, pages 160--169, 1996.

\bibitem{Equiluz02Epidemic}
V.~M. Equiluz and K.~Klemm.
\newblock Epidemic threshold in structured scale-free networks.
\newblock {\em arXiv:cond-mat/02055439}, May 21 2002.

\bibitem{faloutsos99powerlaw}
M.~Faloutsos, P.~Faloutsos, and C.~Faloutsos.
\newblock On power-law relationships of the internet topology.
\newblock In {\em {SIGCOMM}}, pages 251--262, 1999.

\bibitem{GlanceHNSST05}
N.~S. Glance, M.~Hurst, K.~Nigam, M.~Siegler, R.~Stockton, and T.~Tomokiyo.
\newblock Deriving marketing intelligence from online discussion.
\newblock In {\em KDD}, 2005.

\bibitem{Goldenberg:2001}
J.~Goldenberg, B.~Libai, and E.~Muller.
\newblock Talk of the network: A complex systems look at the underlying process
  of word-of-mouth.
\newblock {\em Marketing Letters}, 2001.

\bibitem{Granovetter:1978}
M.~Granovetter.
\newblock Threshold models of collective behavior.
\newblock {\em Am. Journal of Sociology}, 83(6):1420--1443, 1978.

\bibitem{Gruhl:2004a}
D.~Gruhl, R.~Guha, D.~Liben-Nowell, and A.~Tomkins.
\newblock Information diffusion through blogspace.
\newblock In {\em WWW '04}, 2004.

\bibitem{Hethcote:2000}
H.~W. Hethcote.
\newblock The mathematics of infectious diseases.
\newblock {\em SIAM Rev.}, 42(4):599--653, 2000.

\bibitem{Kempe:2003}
D.~Kempe, J.~Kleinberg, and E.~Tardos.
\newblock Maximizing the spread of influence through a social network.
\newblock In {\em KDD '03}, 2003.

\bibitem{Kumar:2003}
R.~Kumar, J.~Novak, P.~Raghavan, and A.~Tomkins.
\newblock On the bursty evolution of blogspace.
\newblock In {\em WWW '03}, pages 568--576. ACM Press, 2003.

\bibitem{jure06viral}
J.~Leskovec, L.~A. Adamic, and B.~A. Huberman.
\newblock The dynamics of viral marketing.
\newblock In {\em EC '06: Proceedings of the 7th ACM conference on Electronic
  commerce}, pages 228--237, New York, NY, USA, 2006. ACM Press.

\bibitem{jurij05patterns}
J.~Leskovec, A.~Singh, and J.~Kleinberg.
\newblock Patterns of influence in a recommendation network.
\newblock In {\em Pacific-Asia Conference on Knowledge Discovery and Data
  Mining (PAKDD)}, 2006.

\bibitem{Richardson:2002}
M.~Richardson and P.~Domingos.
\newblock Mining knowledge-sharing sites for viral marketing, 2002.

\bibitem{Vazquez:2006}
A.~Vazquez, J.~G. Oliveira, Z.~Dezso, K.~I. Goh, I.~Kondor, and A.~L. Barabasi.
\newblock Modeling bursts and heavy tails in human dynamics.
\newblock {\em Physical Review E}, 73:036127, 2006.

\bibitem{Wang02Data}
M.~Wang, T.~Madhyastha, N.~H. Chang, S.~Papadimitriou, and C.~Faloutsos.
\newblock Data mining meets performance evaluation: Fast algorithms for
  modeling bursty traffic.
\newblock {\em ICDE}, Feb. 2002.

\bibitem{WangCWF03}
Y.~Wang, D.~Chakrabarti, C.~Wang, and C.~Faloutsos.
\newblock Epidemic spreading in real networks: An eigenvalue viewpoint.
\newblock In {\em SRDS}, pages 25--34, 2003.

\bibitem{Watts:2002}
D.~J. Watts.
\newblock A simple model of global cascades on random networks.
\newblock In {\em PNAS}, 2002.

\bibitem{Zipf49Human}
G.~Zipf.
\newblock {\em Human Behavior and Principle of Least Effort: An Introduction to
  Human Ecology}.
\newblock Addison Wesley, Cambridge, Massachusetts, 1949.

\end{thebibliography}

\end{document}